\documentclass[preprint]{aastex62}
\usepackage{amsmath,amssymb,CJK,soul}
\usepackage{longtable}
\usepackage{longtable}
\usepackage{threeparttablex} 

\newcommand{\ld}{{P/2019 LD$_2$ }}
\newcommand{\ldns}{{P/2019 LD$_2$}}

\newcommand{\afroz}{{$\mathrm{A}(0^\circ)f\rho$ }}
\newcommand{\afrozns}{{$\mathrm{A}(0^\circ)f\rho$}}

\newcommand{\RomanNumeralCaps}[1]
    {\MakeUppercase{\romannumeral #1}}

\received{--}
\revised{--}
\accepted{--}
\submitjournal{ApJL}

\shorttitle{Initial Characterization of P/2019 LD$_2$ (ATLAS)}
\shortauthors{Bolin et al.}

\begin{document}

\title{Initial Characterization of Active Transitioning Centaur, P/2019 LD$_2$ (ATLAS), using \textit{Hubble}, \textit{Spitzer}, ZTF, Keck, APO and GROWTH Visible $\&$ Infrared Imaging and Spectroscopy}

\correspondingauthor{Bryce Bolin}
\email{bbolin@caltech.edu}

\author[0000-0002-4950-6323]{Bryce T. Bolin}
\affiliation{Division of Physics, Mathematics and Astronomy, California Institute of Technology, Pasadena, CA 91125, USA}
\affiliation{IPAC, Mail Code 100-22, Caltech, 1200 E. California Blvd., Pasadena, CA 91125, USA}

\author{Yanga R. Fernandez}
\affiliation{Department of Physics, University of Central Florida, Orlando, FL 32816, USA}

\author{Carey M. Lisse}
\affiliation{Johns Hopkins University Applied Physics Laboratory, Laurel, MD 20723}

\author[0000-0003-0437-3296]{Timothy R. Holt}
\affiliation{Centre for Astrophysics, University of Southern Queensland, West Street, Toowoomba, QLD 4350, Australia}
\affiliation{Southwest Research Institute, Department of Space Studies, Boulder, CO-80302, USA}

\author{Zhong-Yi Lin}
\affiliation{Graduate Institute of Astronomy, National Central University, 32001, Taiwan}

\author{Josiah N. Purdum}
\affiliation{Department of Astronomy, San Diego State University, 5500 Campanile Dr, San Diego, CA 92182, U.S.A.}

\author[0000-0001-5253-3480]{Kunal P. Deshmukh}
\affiliation{Department of Metallurgical Engineering and Materials Science, Indian Institute of Technology Bombay, Powai, Mumbai-400076, India}


\author{James M. Bauer}
\affil{Department of Astronomy, University of Maryland, College Park, MD 20742-0001, USA}

\author[0000-0001-8018-5348]{Eric C. Bellm}
\affiliation{DIRAC Institute, Department of Astronomy, University of Washington, 3910 15th Avenue NE, Seattle, WA 98195, USA}

\author[0000-0002-2668-7248]{Dennis Bodewits}
\affiliation{Physics Department, Leach Science Center, Auburn University, Auburn, AL 36832, U.S.A.}

\author{Kevin B. Burdge}
\affiliation{Division of Physics, Mathematics and Astronomy, California Institute of Technology, Pasadena, CA 91125, U.S.A.}

\author{Sean J. Carey}
\affiliation{IPAC, Mail Code 100-22, Caltech, 1200 E. California Blvd., Pasadena, CA 91125, USA}

\author{Chris M. Copperwheat}
\affiliation{Astrophysics Research Institute Liverpool John Moores University, 146 Brownlow Hill, Liverpool L3 5RF, United Kingdom}

\author{George Helou}
\affiliation{IPAC, Mail Code 100-22, Caltech, 1200 E. California Blvd., Pasadena, CA 91125, USA}

\author{Anna Y. Q. Ho}
\affiliation{Division of Physics, Mathematics and Astronomy, California Institute of Technology, Pasadena, CA 91125, USA}

\author[0000-0002-1160-7970]{Jonathan Horner}
\affiliation{Centre for Astrophysics, University of Southern Queensland, West Street, Toowoomba, QLD 4350, Australia}

\author{Jan van Roestel}
\affiliation{Division of Physics, Mathematics and Astronomy, California Institute of Technology, Pasadena, CA 91125, U.S.A.}

\author{Varun Bhalerao}
\affiliation{Department of Physics, Indian Institute of Technology Bombay, Powai, Mumbai-400076, India}

\author[0000-0003-1656-4540]{Chan-Kao Chang}
\affiliation{Graduate Institute of Astronomy, National Central University, 32001, Taiwan}

\author{Christine Chen}
\affiliation{Space Telescope Science Institute (STScI), 3700 San Martin Drive, Baltimore, MD 21218, USA}
\affiliation{Department of Physics and Astronomy, The Johns Hopkins University, 3701 San Martin Drive, Baltimore, MD 21218, USA}

\author{Chen-Yen Hsu}
\affiliation{Graduate Institute of Astronomy, National Central University, 32001, Taiwan}

\author{Wing-Huen Ip}
\affiliation{Graduate Institute of Astronomy, National Central University, 32001, Taiwan}

\author{Mansi M. Kasliwal}
\affiliation{Division of Physics, Mathematics and Astronomy, California Institute of Technology, Pasadena, CA 91125, U.S.A.}

\author{Frank J. Masci}
\affiliation{IPAC, Mail Code 100-22, Caltech, 1200 E. California Blvd., Pasadena, CA 91125, USA}

\author[0000-0001-8771-7554]{Chow-Choong Ngeow}
\affiliation{Graduate Institute of Astronomy, National Central University, 32001, Taiwan}


\author{Robert Quimby}
\affiliation{Department of Astronomy, San Diego State University, 5500 Campanile Dr, San Diego, CA 92182, U.S.A.}
\affiliation{Kavli Institute for the Physics and Mathematics of the Universe (WPI), The University of Tokyo Institutes for Advanced Study, The University of Tokyo, Kashiwa, Chiba 277-8583, Japan}

\author{Rick Burruss}
\affiliation{Caltech Optical Observatories, California Institute of Technology, Pasadena, CA 91125, U.S.A.}

\author{Michael Coughlin}
\affil{School of Physics and Astronomy, University of Minnesota, Minneapolis, Minnesota 55455, 
USA}

\author[0000-0002-5884-7867]{Richard Dekany}
\affiliation{Caltech Optical Observatories, California Institute of Technology, Pasadena, CA 91125, U.S.A.}

\author{Alexandre Delacroix}
\affiliation{Caltech Optical Observatories, California Institute of Technology, Pasadena, CA 91125, U.S.A.}

\author{Andrew Drake}
\affiliation{Division of Physics, Mathematics and Astronomy, California Institute of Technology, Pasadena, CA 91125, USA}

\author[0000-0001-5060-8733]{Dmitry A. Duev}
\affiliation{Division of Physics, Mathematics and Astronomy, California Institute of Technology, Pasadena, CA 91125, USA}

\author{Matthew Graham}
\affiliation{Division of Physics, Mathematics and Astronomy, California Institute of Technology, Pasadena, CA 91125, U.S.A.}

\author{David Hale}
\affiliation{Caltech Optical Observatories, California Institute of Technology, Pasadena, CA 91125, U.S.A.}

\author[0000-0002-6540-1484]{Thomas Kupfer}
\affiliation{Kavli Institute for Theoretical Physics, University of California, Santa Barbara, CA 93106, U.S.A.}

\author[0000-0003-2451-5482]{Russ R. Laher}
\affiliation{IPAC, Mail Code 100-22, Caltech, 1200 E. California Blvd., Pasadena, CA 91125, USA}

\author[0000-0003-2242-0244]{Ashish Mahabal}
\affiliation{Division of Physics, Mathematics and Astronomy, California Institute of Technology, Pasadena, CA 91125, U.S.A.}
\affiliation{Center for Data Driven Discovery, California Institute of Technology, Pasadena, CA 91125, U.S.A.}

\author{Przemyslaw J. Mr\'{o}z}
\affiliation{Division of Physics, Mathematics and Astronomy, California Institute of Technology, Pasadena, CA 91125, U.S.A.}

\author{James D. Neill}
\affiliation{Division of Physics, Mathematics and Astronomy, California Institute of Technology, Pasadena, CA 91125, U.S.A.}

\author[0000-0002-0387-370X]{Reed Riddle}
\affiliation{Caltech Optical Observatories, California Institute of Technology, Pasadena, CA 91125}

\author{Hector Rodriguez}
\affiliation{Caltech Optical Observatories, California Institute of Technology, Pasadena, CA 91125, U.S.A}

\author[0000-0001-7062-9726]{Roger M. Smith}
\affiliation{Caltech Optical Observatories, California Institute of Technology, Pasadena, CA 91125}

\author[0000-0001-6753-1488]{Maayane T. Soumagnac}
\affiliation{Lawrence Berkeley National Laboratory, 1 Cyclotron Road, Berkeley, CA 94720, U.S.A.}
\affiliation{Department of Particle Physics and Astrophysics, Weizmann Institute of Science, Rehovot 76100, Israel}

\author{Richard Walters}
\affiliation{Division of Physics, Mathematics and Astronomy, California Institute of Technology, Pasadena, CA 91125, U.S.A.}

\author[0000-0003-1710-9339]{Lin Yan}
\affil{The Caltech Optical Observatories, California Institute of Technology, Pasadena, CA 91125, USA}

\author{Jeffry Zolkower}
\affiliation{Caltech Optical Observatories, California Institute of Technology, Pasadena, CA 91125, U.S.A.}





\begin{abstract}
We present visible and mid-infrared imagery and photometry of temporary Jovian co-orbital comet P/2019 LD$_2$ taken with HST/WFC3, Spitzer/IRAC, the GROWTH telescope network, visible spectroscopy from Keck/LRIS and archival ZTF observations taken between 2019 April and 2020 August. Our observations indicate that the nucleus of LD$_2$ has a radius between 0.2-1.8 km assuming a 0.08 albedo and a coma dominated by $\sim$100$\mu$ m-scale dust ejected at $\sim$1 m/s speeds with a $\sim$1\arcsec~jet pointing in the SW direction. LD$_2$ experienced a total dust mass loss of $\sim$10$^8$ kg at a loss rate of $\sim$6 kg/s with Af$\rho$/cross-section varying between $\sim$85 cm/125 km$^2$ and $\sim$200 cm/310 km$^2$ from 2019 April 9 to 2019 Nov 8. If the increase in Af$\rho$/cross-section remained constant, it implies LD$_2$'s activity began $\sim$2018 November when within 4.8 au of the Sun, implying the onset of H$_2$O sublimation. We measure CO/CO$_2$ gas production of $\lesssim$10$^{27}$ mol/s /$\lesssim$10$^{26}$ mol/s from our 4.5 $\mu$m Spitzer observations, $g$-$r$ = 0.59$\pm$0.03, $r$-$i$ = 0.18$\pm$0.05, $i$-$z$ = 0.01$\pm$0.07 from GROWTH observations, H$_2$O gas production of $\lesssim$80 kg/s scaling from our estimated $C_2$ production of  $Q_{C_2}\lesssim$7.5$\times10^{24}$ mol/s from Keck/LRIS spectroscopy. We determine that the long-term orbit of LD$_2$ is similar to Jupiter family comets having close encounters with Jupiter within $\sim$0.5 Hill radius in the last $\sim$3 y, within 0.8 Hill radius in $\sim$9 y. Additionally, 78.8$\%$ of our orbital clones are ejected from the Solar System within $1 \times 10^{6}$ years having a dynamical half-life of 3.4 $\times 10^5$ years.
\end{abstract}
\keywords{minor planets, asteroids: individual (\ldns), temporarily captured orbiters, minimoons}

\section{Introduction}
The gas giant Jupiter is the dominant gravitational perturbing body affecting the dynamical transfer of Solar System comets from the outer Solar System's trans-Neptunian disk beyond the orbit of Neptune into the inner reaches of the Solar System \citep[recently described in][]{Dones2015}. The vast majority of comets in transfer from the outer Solar system regions such as the Oort Cloud in the case of long period comets \citep[recently described in][]{Vokrouhlicky2019} or the trans-Neptunian region in the case of short period comets \citep[recently described in][]{Nesvorny2017}. Once the comets originating from the trans-Neptunian region randomly walk their way through the outer Solar System and become strongly influenced by close-encounters with Neptune and Uranus, a significant portion are transformed in their orbital configuration into the Centaur group of small bodies. The Centaur class is defined as having semi-major axes, $a$, and perihelion, $q$, between 5.2 au, the semi-major axis, $a_J$ of Jupiter, and 30 au, the semi-major axis of Neptune, $a_N$, \citep[][]{Jewitt2009}. An additional quantity used to define the small bodies in the inner-Solar system is the Tisserand parameter with respect to Jupiter, $T_J$, defined as~\begin{equation}T_J \, = \, \frac{a_J}{a} \, + \, 2\,\sqrt{(1-e^2)\,\frac{a}{a_J}} \, \mathrm{cos}i\end{equation}~where $e$ is the eccentricity of the body and $i$ is the inclination. $T_J$ can be used as a rough indication of how much an object is influenced by the gravitational perturbations of Jupiter \citep[][]{Murray1999}. Centaurs are objects that generally have $T_j>$3.05 \citep[][]{Gladman2008} whereas Jupiter family comets have 3$>T_J>$2\citep[][]{Duncan2004}. However, we note that the $T_j>$3.05 boundary does not strictly define objects as Centaurs as there can be non-Centaur objects with $T_j>$3.05.

The mean dynamical half-life of Centaurs is $\sim$2.7 Myr with the vast majority of Centaurs eventually being ejected from the Solar System \citep[][]{Horner2004CentaursI} while the Jupiter family comets have a bit shorter lifetimes of $\sim$0.5 Myr \citep[][]{Levison1994}. The chaotic evolution of the Centaurs causes a significant number (around one third) to becoming Jupiter family comets at some point in their lifetimes, prior to their eventual ejection from the Solar System \citep[][]{Horner2004b}. The Centaur 2014 OG$_392$, recently discovered to be active \citep[][]{Chandler2020}, may be an example of an object transitioning between the Centaur and Jupiter family comet groups. Some can even be temporarily captured as satellites of the giant planets, or to the Jovian and Neptunian Trojan populations \citep[e.g.,][]{Horner2006,Horner2012}. Another example of a Centaur recently in the stage of becoming a Jupiter Family Comet is 29P/Schwassmann-Wachmann. Centaur 29P is located in a region of orbital parameters space with 5.5 au $<q<$8.0 au, and aphelion, 5 au $<Q<$ 7 au that acts as a ``gateway'' that the Centaurs preferentially inhabit whilst in the process of dynamically transferring to become Jupiter family comets to becoming Jupiter family comets \citep[][]{Sarid2019}.

The recently discovered briefly Jovian co-orbital comet \ld \citep[][]{Sato2020CBET}, with a semi-major axis of 5.30 au, a perihelion of 4.57 au and aphelion of 6.02, may be another example of an object in the transition region between Centaur objects and Jupiter family comets. The comet will only spend $\sim$1 orbit in the dynamical configuration where it has a Jupiter-similar semi-major axis \citep[][]{Hsieh20202019ld2, Kareta2020RNAAS}. Initially reported as an inactive object by the ATLAS survey \citep[][]{Tonry2011} in 2019 June and designated by the Minor Planet Center as 2019 LD$_2$\footnote{\texttt{https://minorplanetcenter.net/db$\_$search/show$\_$object?utf8=$\%$E2$\%$9C$\%$93$\&$object$\_$id=P$\%$2F2019+LD2}}, it was discovered to be active by amateur astronomers\footnote{\texttt{http://aerith.net/comet/catalog/2019LD2/2019LD2.html}}. Pre-discovery images and follow-up images of the comet taken by ATLAS and other ground-based telescopes resulted in it being given the cometary designation \ld \citep[][]{Fitzsimmons2020MPEC}. Whilst technically some of the orbital elements of \ld such as its semi-major axis, resemble those of a Jovian co-orbital, it is inherently unstable in stark contrast to the stable orbits of Jovian Trojans, which are stable on timescales comparable to the age of the Solar system and are located at $\sim\pm$60$^\circ$ mean longitude with respect to Jupiter \citep[e.g.,][]{Marzari2002}. In addition, the Jovian Trojans have a different origin having most likely been captured as a result of Jupiter's migration during the Solar system's formation, 4.5 Gyr ago \citep[e.g.,][]{Morbidelli2005,Roig2015}.

One proposed origin for \ld is that it is a Jupiter family comet in the transition region in orbital parameters space inhabited by objects that are in transition between Centaurs and Jupiter family comets \citep[][]{Steckloff2020}.
As comets transfer from their origins in the outer Solar System beyond the orbit of Neptune and become denizens of the inner Solar System, they will experience a dramatic shift in the thermal environment due to increased thermal insolation from the Sun \citep[][]{DeSanctis2000,Sarid2009}. The consequence of the increased Solar insolation as the comet nears the Sun is the increased heating and sublimation of volatiles such as CO and H$_2$O near the comet's surface \citep[][]{Meech2004,Lisse2020}. Another consequence of the increased heating from closer proximity of the Sun is that large-scale ablation of the comet's structure due to thermal stress can occur resulting in it becoming partially or completely disrupted \citep[][]{Fernandez2009}. Since \ld is now in transition between the Centaur and Jupiter Family Comet populations, it seems likely that it has become active for the first time, and as such, its activity will be rapidly evolving in response to the new epoch of increased Solar heating.

We therefore present in this paper the analysis of visible light high-resolution \textit{Hubble Space Telescope}/Wide Field Camera 3 \citep[\textit{HST}/WFC3][]{Dressel2012} observations of \ld using the approach of \citet[][]{Jewitt2014} and \citet[][]{Bolin2020HST} to understand the dust coma and nucleus properties, and to constrain the cause of \ldns's activity. We will also use mid-infrared (MIR) \ld observations taken with \textit{Spitzer Space Telescope}/Infrared Array Camera \citep[\textit{Spitzer}/IRAC][]{Werner2004} combined with the analysis techniques of \citet[][]{Reach2013} and \citet[][]{Lisse2020Spitzer} to place upper limits on the comet's CO+CO$_2$ gas production. We also use multi-wavelength observations covering the visible and MIR by building on the techniques of \citet[][]{Bolin2020asdfasdf} by using a network of ground-based observatories to characterize the physical properties of this transitioning Centaur. In addition, we will examine the long-term orbital properties of \ld using its latest orbital solution in order to better understand its possible origins and future dynamical evolution.

\section{Observations}
\label{s.obs}
Observations of \ld were obtained before the official announcement of its activity in 2020 May both by targeted observations by ground and space-based observatories and serendipitously in the survey observations by the Zwicky Transient Facility (ZTF) \citep[][]{Graham2019}. The time span of our targeted observations is 2019 Sep 7 UTC to 2020 Aug 19 UTC including observations by the Astrophysical Research Consortium 3.5 m telescope (ARC 3.5 m) \textit{Spitzer}, \textit{HST}, Keck I and members of the GROWTH network \citep[][]{Kasliwal2019} such as Mount Laguna Observatory 40-inch Telescope (MLO 1.0-m), Liverpool Telescope (LT), Lulin Optical Telescope (LOT). A list of our targeted observations and their viewing geometry is listed in Table~\ref{tab:obs}. The time span of our serendipitous observations of \ld made with the ZTF survey is between 2019 April 9 UTC and 2019 Nov 8 UTC and are listed with their viewing geometry in Table~\ref{tab:phot}.

\subsection{Zwicky Transient Facility}
We searched for serendipitous observations of \ld made with the Zwicky Transient Facility survey mounted on the Palomar Observatory's 48-inch telescope \citep[][]{Bellm2019} in the ZTF archive \citep[][]{Masci2019}. The ZTF archive possessed observations of \ld made as far back as 2019 April 9 UTC which we include up to 2019 Nov 8 UTC. The observations were made in $g$ and $r$ band in images consisting of 30 s exposures. Seeing conditions were typically between 1.5-2.5\arcsec~ and at air masses ranging from 1.4 to 2.6. A full list of observations of \ld made by ZTF containing the viewing geometry and observing conditions is presented in Table~\ref{tab:phot}.
\subsection{Apache Point Astrophysical Research Consortium 3.5 m}Following the announcement of the appearance of activity of \ldns$^2$ (then called 2019 LD$_2$), we triggered target of opportunity observations with the ARC 3.5 m at Apache Point Observatory on 2019 September 7 UTC using the ARCTIC large-format optical CCD camera \citep[][]{Huehnerhoff2016}. The camera was used in full-frame, quad amplifier readout, 2$\times$2 binning mode resulting in a pixel scale of 0.228\arcsec~ and used with the $g$ and $r$ filters. In total, 14 $g$ and $r$ exposures were obtained, each 120 s long and in alternating order between the $g$ and $r$ filters. The telescope was tracked at the sky-motion rate of the comet of 8.6\arcsec/h. The seeing was 1.4\arcsec~ and the airmass was 1.8 during the observations.

\subsection{\textit{Spitzer Space Telescope}}Observations of \ld were made with the \textit{Spitzer Space Telescope} (\textit{Spitzer}) using the IRAC instrument \citep[][]{Fazio2004} on 2020 January 25-26 UTC \citep[DDT program 14331, PI][]{Bolin2019Spitzer}. The observations consisted of 11 Astronomical Observing Requests (AORs), each consisting of 80 x 12 s dithered frames and having a $\sim$0.44 h duration for a total of 4.8 h clock time. The frames where dithered in groups of 10, with each using a large cycling pattern. The sky at the location of \ld during the \textit{Spitzer} observations possessed a high density of stars due to its low, -18$^\circ$ galactic latitude, therefore shadow observations were used to improve the sensitivity of the observations. Out of a total of 11 AORs, eight were focused on observed \ld for a total of 2.13 h on source time. The remaining three AORs were Shadow observations that were evenly spaced in the sky location covering the trajectory of \ld between 2020 January 02:23:32-23:10:44 that \ld was being observed. The target was centered in the 4.5 $\mu$m channel since this channel is sensitive to CO/CO$_2$ emission and also because the object was expected to be brightest at this wavelength. The 4.5 $\mu$m IRAC channel has a spatial resolution of 1.2\arcsec/pixel. The data were reduced in a method as described in \citep[][]{Fernandez2013}.
\subsection{\textit{Hubble Space Telescope}}The \textit{Hubble Space Telescope} (\textit{HST}) was used to observe \ld with General Observer's (GO) time on 2020 April 1 UTC \citep[HST GO 16077, PI][]{BolinHST2020}. During the one orbit visit, five 380 s F350LP filter exposures were obtained with the UVIS2 array of the WFC3/UVIS camera \citep[][]{Dressel2012} for a total of 1900 s integration time over a single orbit. The F350LP filter has a central wavelength of 582 nm with a FWHM bandpass of 490 nm \citep[][]{Deustua2017}. The instrument and filter combination of WFC3 and the F350LP filter provides a per-pixel resolution of 0.04\arcsec~ corresponding to 145 km at the topo-centric distance of the comet. The comet was tracked non-sidereally according to its skyplane rate of motion of 40\arcsec/h. 
\subsection{Mount Laguna Observatory 40-inch Telescope}
Multi-band optical images of \ld were obtained with the 1.0 m Telescope at the Mount Laguna Observatory \citep[][]{Smith_Nelson1969} on 2020 May 17 UTC. Johnson-Cousins $B$, $V$ and $R$ filters were used in combination with the E2V 42-40 CCD Camera to obtain 7-9 120 s exposures in each filter. The seeing conditions were 1.92\arcsec, the airmass was 1.49 and sidereal tracking was used. This facility is a member of the GROWTH collaboration.
\subsection{Liverpool Telescope}
Observations of \ld were made in $g$, $r$, $i$ and $z$ filters by the 2 m Liverpool Telescope located at the Observatorio del Roque de los Muchachos on 2020 May 29 UTC. The IO:O wide-field camera was used with a 2x2 binning providing a pixel scale of 0.3\arcsec~ \citep[][]{Steele2004}. Two 30 s exposures were made per filter with the telescope tracking at the sidereal rate. The seeing conditions were 1.21 \arcsec~ and the airmass was 1.75. Detrending of data was performed using the automated IO:O pipeline software \citep[][]{Steele2004}. This facility is a member of the GROWTH collaboration.
\subsection{Lulin Optical Telescope}
Multiband $B$, $V$, and $R$ imaging of \ld were made by the 1 m Lulin Optical Telescope on 2020 June 23-27 UTC and 2020 July 10 UTC. The observations were made using the 2K $\times$ 2K SOPHIA camera with a pixel scale of 0.52 \arcsec \citep[][]{Kinoshita2005}. Exposure times of 90 s were where the telescope was tracked at the non-sideral rate determined by the ephemeris of the comet. The seeing conditions of the observations were $\sim$1.5\arcsec~ and the airmass was $\sim$1.15.
\subsection{Keck I Telescope}
 A spectrum of \ld was obtained using the Low-Resolution Imaging Spectrometer (LRIS) \cite{Oke1995} on the Keck I telescope on 2020 August 19 UTC (PI J. von Roestel, C272). The blue camera consisting of a 2 x 2K x 4K Marconi CCD array was used with the red camera consisting of a science grade Lawrence Berkeley National Laboratory 2K x 4K CCD array. Both cameras have a spatial resolution of 0.135 \arcsec/pixel. We used the 560 nm dichroic with $\sim$50$\%$ transmission efficiency in combination with the 600/4000 grism for the blue camera, rebinned twice in both the spectral and spatial direction, and the 600/7500 grating for the red camera, rebinned once in the spectral direction and twice in the spatial direction, providing a spectral resolution of 0.8 nm and 0.5 nm, respectively, and a spatial resolution of 0.27\arcsec. A total integration time of 300 s was used for the exposure and was obtained at airmass 1.8 in 0.85\arcsec~ seeing conditions. Both telluric correction and Solar-analog stars were observed at similar air masses as \ldns. Wavelength calibration was completed using the HgCdZn lamps for the blue camera and the ArNeXe lamps for the red camera. We used a local Solar analog star to remove the Solar component from the spectrum of \ldns. The LPipe spectroscopy reduction package was used to reduce the data \citep[][]{Perley2019}.
 
\section{Results}
\subsection{Morphology and nucleus}
\label{sec:morph}
 
\begin{figure}
\centering
\includegraphics[scale=.55]{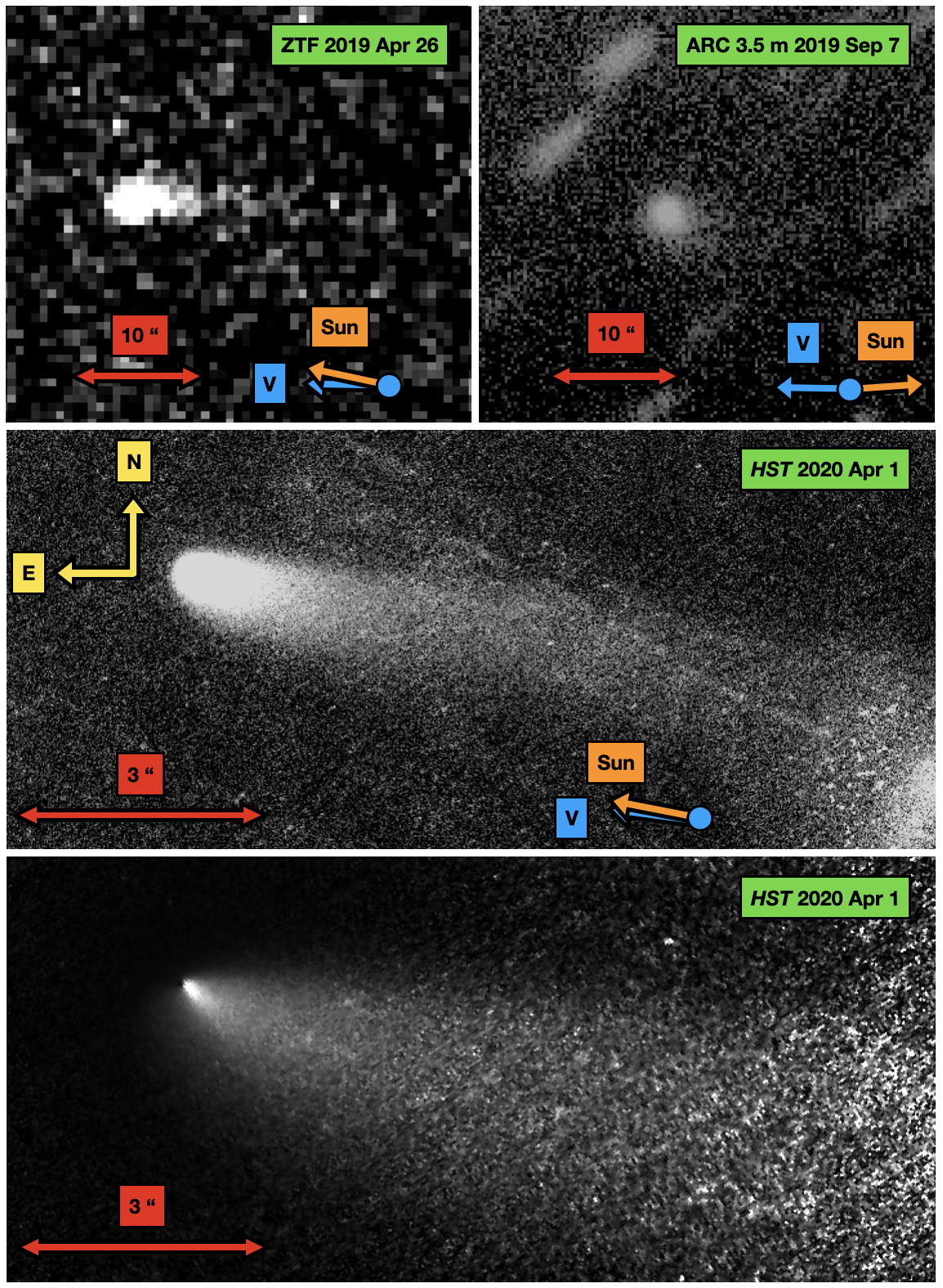}
\caption{Top left panel: a 90 s equivalent exposure time stack of 3 x 30 s $r$ filter images of \ld taken by ZTF on 2019 April 26 UTC. The image stack was compiled using the ZChecker software \citep[][]{Kelley2019}. The pixel scale is 1\arcsec~/pixel and the seeing was $\sim$2.2\arcsec. An arrow indicating the width of 10\arcsec~is shown for scale, equivalent to $\sim$30,000 km at the geocentric distance of 4.15 au of the comet on 2020 April 26 UTC. The solar, orbital velocity and cardinal directions are indicated. Top right panel: a 2,400 s equivalent exposure time robust mean stack of 20 x 120 s $r$ filter images of \ld taken with the ARC 3.5 m on 2019 September 7 UTC. The telescope was tracked at the comet's motion. The pixel scale is 0.228\arcsec/pixel and the seeing was $\sim$1.4\arcsec. An arrow indicating the width of 10\arcsec~is shown for scale, equivalent to $\sim$31,000 km at the geocentric distance of 4.28 au of the comet on 2020 September 7 UTC. Center panel: a 1,900 s equivalent exposure time robust mean stack of 5 x 380 s F350LP filter images of \ld taken with \textit{HST}/WFC3 on 2020 April 1 UTC. The pixel scale is 0.04\arcsec/pixel. An arrow indicating the width of 3\arcsec~is shown for scale, equivalent to $\sim$11,000 km at the geocentric distance of 5.02 au of the comet on 2020 April 1 UTC. Bottom panel: the same as the center panel but normalized according to the radial profile of the comet. A $\sim$1\arcsec~jet-like structure is seen with a position angle of $\sim$210$^\circ$.}
\label{fFiig:mosaic}
\end{figure}

 Serendipitous pre-discovery observations of \ld were obtained with ZTF on 2019 April 26 UTC consisting of three 30 s exposures in $r$ band. These pre-discovery data of \ld have been co-added into a composite image with an equivalent 90 s integration time presented in the top left panel of Fig.~\ref{fFiig:mosaic}. The comet has an extended appearance with a $\sim$20\arcsec~ long tail with a position angle of $\sim$260$^\circ$ in the anti-Solar direction. ZTF obtained pre-discovery detections of \ld on 2019 April 9,15 and 20, but the comet did not have a discernible extended appearance in these data.
 
 On 2019 September 7 UTC, the ARC 3.5 m was used to obtain 20 x 120 s exposures of \ld in $r$ band. A composite median stack with an equivalent exposure time of 2400 s is presented in the top right panel of Fig.~\ref{fFiig:mosaic}. In the ARC 3.5 m images, the comet has a diffuse, non-stellar appearance. The tail is not easily defined in the ARC 3.5 m median stack, though the comet's extended appearance is enhanced in the opposite direction of the comet's orbital motion with a position angle of $\sim$230$^\circ$ and length of 5\arcsec.
 
 The center panel of Fig.~\ref{fFiig:mosaic} presents the appearance of \ld in a median stack of five 380 s F350LP image with an equivalent integration time of 1900 s taken with \textit{HST}/WFC3 on 2020 April 01 UTC. Cosmic rays have been removed from the composite image stack with median interpolation of the surrounding pixels. The high-resolution composite \textit{HST} stack was taken when the comet was at an orbit-plane angle of $\sim$-0.44$^\circ$ and had a tail with a length of $\sim$32\arcsec~ limited by background structure caused by galaxies and sky noise opposite of the Solar direction with a position angle of $\sim$250$^\circ$. The $\sim$32\arcsec-long tail translates into a length of 6.2$\times$10$^{8}$ m given its topo-centric distance of 5.02 au and a phase angle of 10.7$^\circ$. An enhanced version of the \textit{HST} median composite stack normalized by the distance from the optocenter reveals a possible jet structure $\sim$1\arcsec~ long as seen in the bottom panel of Fig.~\ref{fFiig:mosaic}. We will discuss the implications of the comet's morphology from these observations for its dust properties below in Section~\ref{sec:dust}.

\begin{figure}
\centering
\includegraphics[scale=.45]{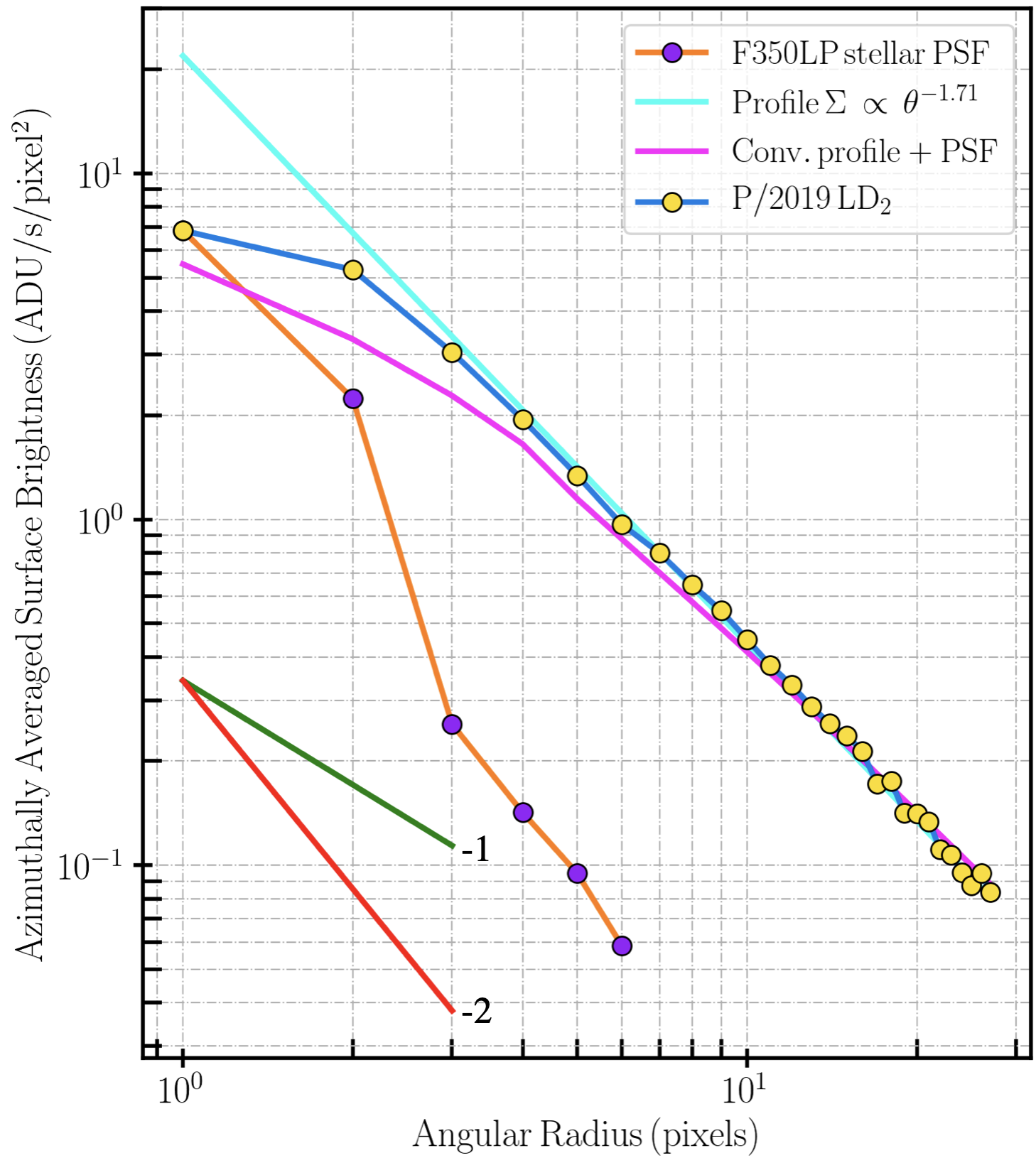}
\caption{The normalized surface brightness profile of \ld taken with \textit{HST}/WFC3 on 2020 April 1 UTC presented as the yellow circles with a connecting blue line. A surface brightness profile of $\Sigma \, \propto \, \theta^{-1.71}$ fitted to the profile of \ld between 0.24\arcsec~ and 1.20\arcsec~ is plotted as the cyan line using the same vertical scale as the the normalized surface brightness profile of \ld. The normalized surface brightness profile of a F350LP stellar PSF assuming a G2V-like source generated using TinyTim \citep[][]{Krist2011} is plotted as purple circles with a connecting orange line. The surface brightness profile resulting from the convolution of the F350LP stellar PSF and the fitted $\Sigma \, \propto \, \theta^{-1.71}$ surface brightness profile of \ld is plotted as a pink line using the same vertical scale as the the normalized surface brightness profile of \ld. Logarithmic Surface brightness gradients with $m$ = -1 and $m$ = -2 are plotted as green and red lines respectively for comparison. Statistical error bars on the surface brightness computed assuming Poissonian statistics at each radius element are smaller than the plot symbols used for both \ld and the synthetic stellar PSF.}
\label{fFiig:radial}
\end{figure}

We compare the surface brightness profile of \ld to the simulated surface brightness profile of a G2 field star WFC3 point-spread function (PSF) assuming the use of the F350LP filter using the TinyTim software \citep[][]{Krist2011} as seen in Fig.~\ref{fFiig:radial}. Both the radial profiles of \ld and the simulated stellar G2V source are computed by azimuthally averaging concentric apertures centered on the optocenter separated by the pixel scale allowed by WFC3 using the F350LP filter.  The normalized surface brightness profile of \ld between 0.24\arcsec-1.2\arcsec~ was fit to the functional form of $\Sigma\,\propto\,\theta^{m}$ where $\Sigma$ is the surface brightness and $\theta$ is the distance from the optocenter in pixels resulting in a radial profile slope of m$\sim$-1.71. We note that the radial profile slope is steeper than the typical -1 to -1.5  radial profile slope of comets with an isotopic coma in a steady state. The steeper radial profile slope of \ld compared to comets with isotopic coma may be an independent indication that the comet's evolving dust production rate  \citep[][]{Jewitt1987ab}.

The fitted 0.24\arcsec-1.2\arcsec~ radial profile of \ld was convolved with the synthetic G2V PSF and subtracted from the measured radial profile of \ld to calculate a equivalent nucleus brightness of V = 22.6$\pm$0.04 assuming a m$_V$-m$_{\mathrm{F350LP}}$$\sim$0.1\citep[][]{Bolin2020asdfasdf}. We assume the following phase function for determining the absolute magnitude of the nucleus, $H$ 
\begin{equation}
\label{eqn.brightness}
H = V - 5\, \mathrm{log_{10}}(r_h \Delta) - \Phi(\alpha)
\end{equation}
where $r_h$ and $\Delta$, $\alpha$ are the heliocentric distance, topo-centric distance and phase angle of the comet as listed in Table~\ref{tab:obs} for the 2020 April 1 UTC observation. $\Phi(\alpha)$ = 0.04$\alpha$, where we assume a phase coefficient of 0.04 in magnitudes/degree, resulting in $H$ = 15.53$\pm$0.05. The true phase coefficient of \ld is unknown, therefore our uncertainty on the measured value of $H$ is considered a lower limit.

From our measured value of $H$, we calculate the light scattering cross-section, $C$, of \ld in km$^2$ using the following function
\begin{equation}
\label{eqn.crosssetion}
C = 1.5 \times 10^{6}\, p_v^{-1} \, 10^{-0.4H}
\end{equation}
where $p_v$ is the albedo the nucleus, assumed to be $\sim$0.08, the typical albedo measured for Centaurs \citep[][]{Bauer2013}, resulting in $C$=11.15$\pm$0.42 km$^2$. Converting our measured cross-section to a radius using $r = (C/\pi)^2$, we obtain a radius of $\sim$1.8 km, comparable to the radius estimates of \ld based on un-resolved photometry and the non-detection of \ld from ground-based observations \citep[][]{Schambeau2020cbet}. We note that this is a radius estimate based on a single observation and represents a size assuming a spheroid shape. Significant deviations from a spheroid shape such as a bi-lobal \citep[][]{Nesvorny2018} or elongated shape \citep[][]{Bolin2018,Hanus2018} as has been observed for other comet-like bodies may require additional observations to be made of \ld to accurately determine its size.

\subsection{Photometry and lightcurve}
\label{sec:phot}
Using the combination of our ground-based observations with the ARC 3.5 m taken on 2019 September 7 UTC, the MLO 1.0-m on 2020 May 27 UTC, the LT on 2020 May 29 UTC and Lulin Optical Observatory on 2020 July 10 UTC, we have calculated the mean colors of \ld using 10,000 km photometric apertures of $g$-$r$ = 0.60$\pm$0.03, $r$-$i$=0.18$\pm$0.05, $i$-$z$=0.01$\pm$0.07. The filter configuration and viewing geometry of our observations are presented in Table~\ref{tab:obs}. The equivalent angular size of the 10,000 km used in our photometric calculations ranged from 3.2\arcsec~ to 3.7\arcsec~ with the seeing during observations ranging from 1.2\arcsec~ to 1.9\arcsec. We used the color transformations from \citet[][]{Jordi2006} to convert the $BVR$ Johnson-Cousins photometry of \ld from the MLO 1.0-m and LT to the SDSS system.

\begin{figure}
\centering
\includegraphics[scale=.52]{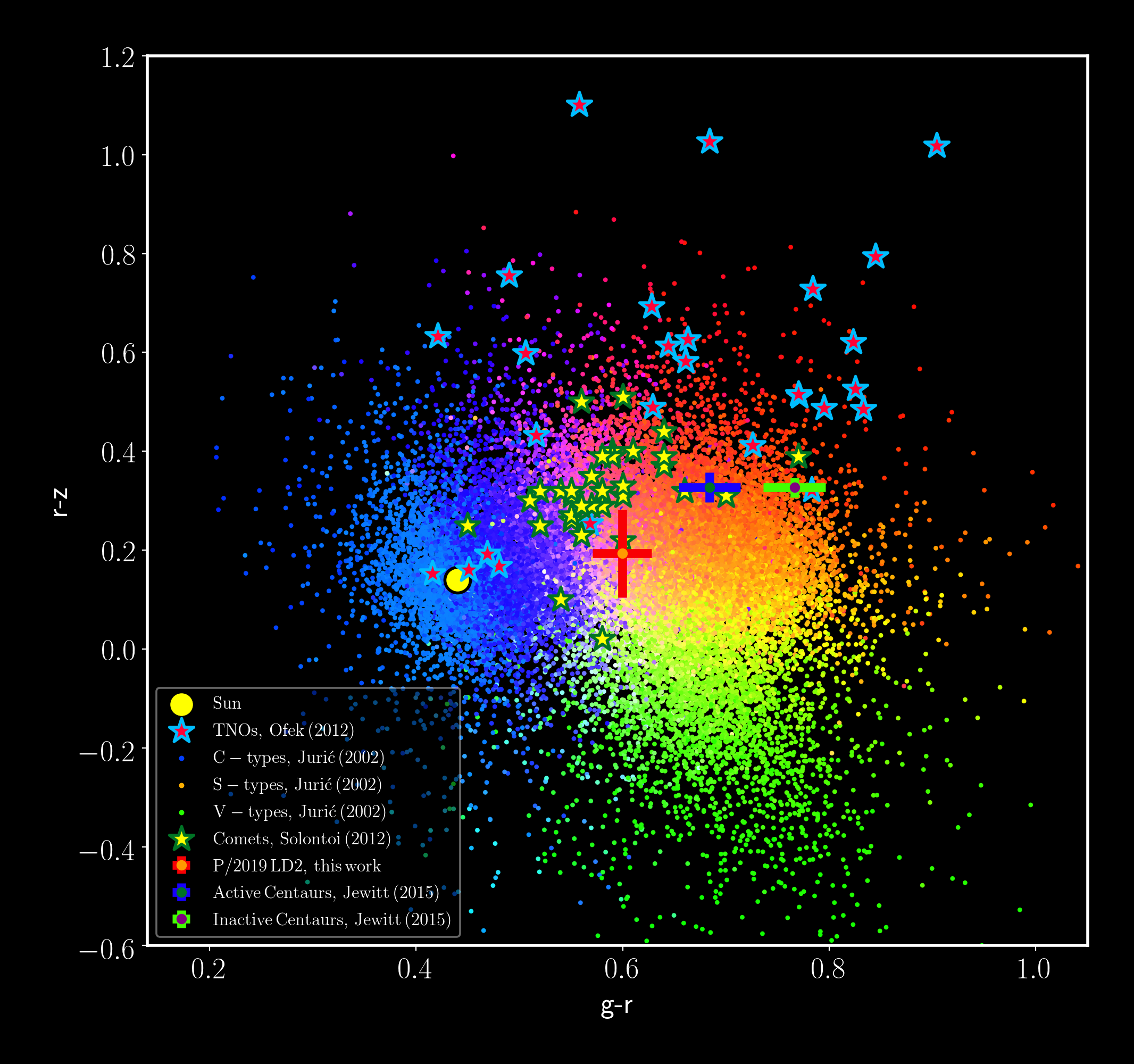}
\caption{$g-r$ vs. $r-z$ colors of \ld plotted with $g-r$ and $r-z$ colors of inactive Solar System C, S and V-types asteroids \citep[][]{Ivezic2001,Juric2002,DeMeo2013}, active comets \citep[][]{Solontoi2012} and Kuiper Belt objects \citep[][]{Ofek2012}. The colorization scheme of data points for asteroids by their $griz$ colors is adapted from \citet[][]{Ivezic2002}. The colors of active and inactive Centaurs from \citep[][]{Jewitt2015bbb} is also included. The most appropriate color comparison between the color of \ld and other Solar System bodies is between the active comets in \citet[][]{Solontoi2012} and the active Centaurs from \citet[][]{Jewitt2015bbb} because the colors of \ld are most representative of its dust rather than bare nucleus. The colors of inactive bodies are included for comparison purposes only.}
\label{fFiig:colors}
\end{figure}

Our visible measured colors of \ld are reddish to neutral in the $\sim$480 nm to $\sim$910 n wavelength range covered by our filters consistent with the measured colors of other active Solar System comets as presented in Fig.~\ref{fFiig:colors}. For comparison purposes only, we have included the colors of inactive objects in Fig~\ref{fFiig:colors}. We note that the measured colors of \ld from our observations are somewhat bluer compared to the colors of active and inactive Centaurs measured by \citet[][]{Jewitt2015bbb}, though this may be due to the longer wavelength coverage of our observations which go as far as $\sim$910 nm compared to the shorter visible-wavelength observations of \citet[][]{Jewitt2015bbb}.

Images from each of the  \textit{Spitzer} DDT program 14331 AORs 1,4,6,9 and 10 that were used to take images of \ld from 2020 January 25 2:23 to 23:11 UTC were reduced using the reduction methods described in \citet[][]{Fernandez2013}. Images obtained during each of these five AORs using the 4.5 $\mu$m channel were co-added to form a single composite image with an equivalent exposure time of 948 s for each AOR. 

\begin{figure}
\centering
\includegraphics[scale=.72]{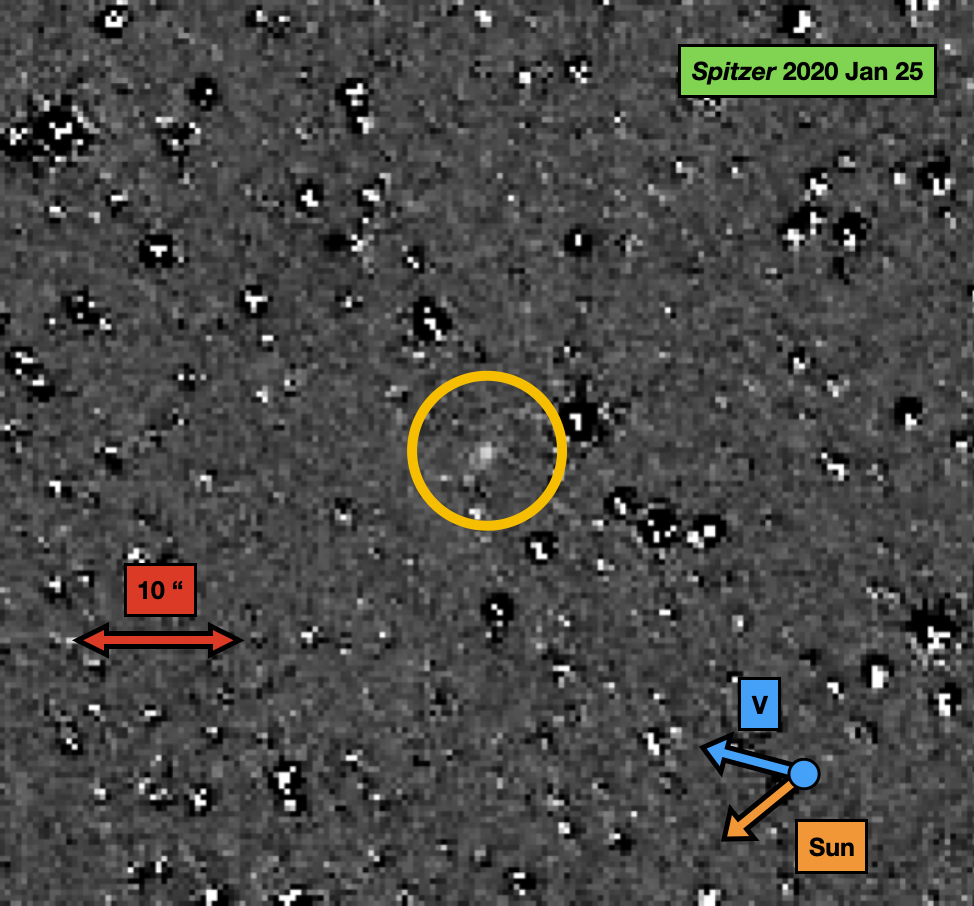}
\caption{A 948 s equivalent integration time composite image stack made from \textit{Spitzer}/IRAC observations of \ld taken during \textit{Spitzer} DDT program 14331 AOR 1 on 2020 January 25 02:23:32 UTC. The detection of \ld has been encircled in yellow. The pixel scale is 1.2\arcsec/pixel. An arrow indicating the width of 10\arcsec~is shown for scale, equivalent to $\sim$31,000 km at the topo-centric distance of 4.256 au of the comet on 2020 January 25 UTC. The solar, orbital velocity and cardinal directions are indicated.}
\label{fFiig:sst}
\end{figure}

\ld was located in a crowded star field at a galactic latitude of $\sim$-18$^\circ$. Therefore, due to the imperfections in the shadowing technique, we used an aperture size with an angular width of 3.24\arcsec~ equivalent to 10,000 km at the topo-centric distance of 4.26 au of the comet from \textit{Spitzer}. We obtain an on-source flux density for \ld of 35.6$\pm$2.8 $\mu$Jy at 4.5 $\mu$m using the average of the five photometry measurements from the composite images made from each of the AORs 1, 4, 6, 9 and 10. The comet has a slightly extended appearance of more than $\sim$2.4 \arcsec~ as seen in Fig.~\ref{fFiig:sst} and an aperture correction was applied to the flux density measurement. The flux from the nucleus assuming a $\sim$km-scale nucleus radius as measured in Section~\ref{sec:morph} is $\sim$0.1 $\mu$Jy, less than 1$\%$ of the total flux.

We present the $\mathrm{A}f\rho$ based off of the \textit{Spitzer}/IRAC photometry of \ld using the $\mathrm{A}f\rho$ definition of \citet[][]{AHearn1984} which is a quantity in units of length, in this case cm, that corrects the comet's brightness with respect to heliocentric distance, geocentric distance, aperture size, Solar spectrum and filter wavelength. The values $\mathrm{A}f\rho$ are normalized to 0$^{\circ}$ phase angle, $A(0^{\circ})f\rho$, using the Halley-Marcus cometary phase function defined by \citet[][]{Schleicher2011}. Assuming that the entirety of the flux in the 4.5 $\mu$m \textit{Spitzer}/IRAC observations is from dust in local thermal equilibrium, we calculate a $A(0^{\circ})f\rho$ = 334 cm. This is strictly an upper limit on $A(0^{\circ})f\rho$ due to the possible contribution of gas in the measured flux of the comet.

If we assume that the entirety of the flux from the comet is from CO emission and that the gas speed is 0.5 km/s, we measure a gas production rate of 1.6$\pm$0.1$\times$10$^{27}$mol/s similar to the results of \citet[][]{Kareta2020ld2}. For CO$_2$, assuming that the entirety of the flux is due to gas emission, we obtain a gas production rate of 1.4$\pm$0.1$\times$10$^{26}$mol/s comparable to the CO$_2$ measured for comets observed in the MIR at similar heliocentric distances \citep[][]{Reach2013,Bauer2015}.

\begin{figure}
\centering
\includegraphics[scale=0.405]{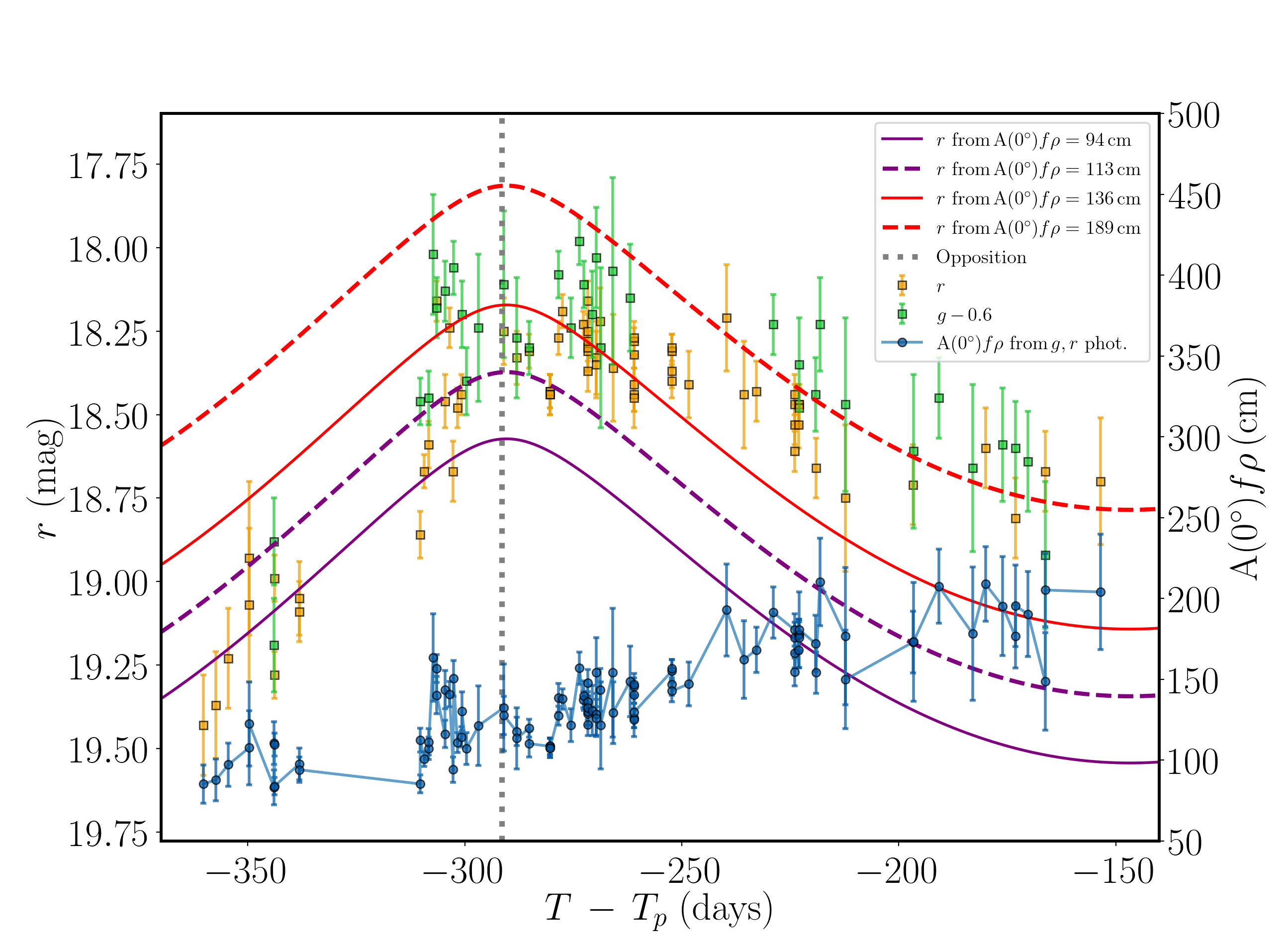}
\includegraphics[scale=0.405]{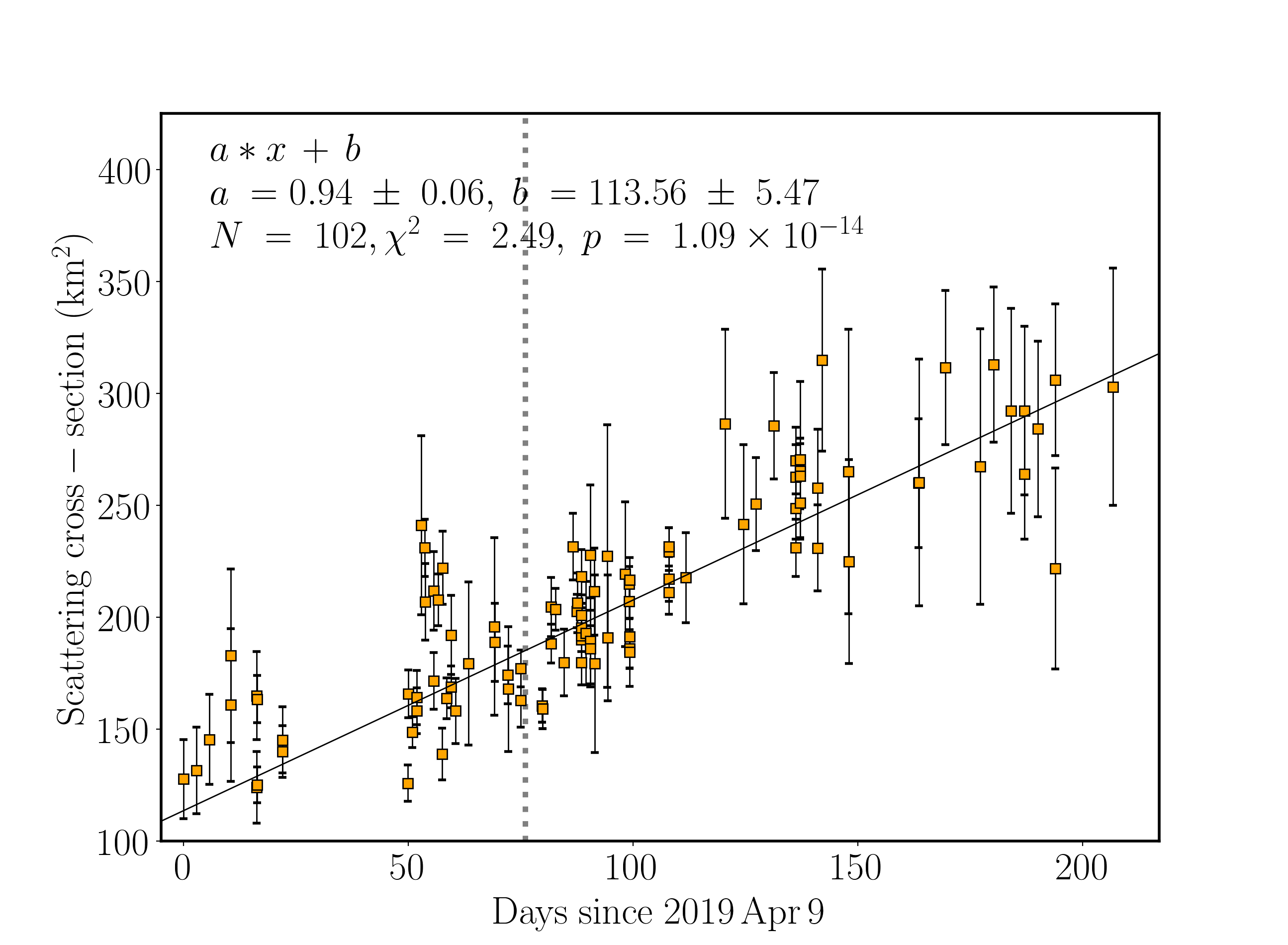}
\caption{Top panel: $g$ and $r$ band photometric lightcurve of \ld vs. time from perihelion ($T - T_p$) measured using a fixed 20,000 km aperture between 2019 April 9 UTC and 2019 November 8 UTC using the data from Table~\ref{tab:phot}. Multiple lines of $A(0^{\circ})f\rho$ =  94 cm, 113 cm, 136 cm, 189 cm are shown as purple and red solid and dashed lines to reflect the change in the comet's brightness as the comet moved through opposition at ($T - T_p$) = -291.5 days on 2019 June 24 UTC which is shown as a vertical grey dotted line. In addition, values of $A(0^{\circ})f\rho$ calculated using the brightness values and viewing geometry in  Table~\ref{tab:phot} are presented as dark blue data points connected by a light blue solid line. Bottom panel: The scattering cross-section of \ld calculated from Eq.~\ref{eqn.crosssetion} as a function of days since 2019 April 09 UTC from the photometric data presented in Table~\ref{tab:phot}. The black line shows the minimized $\chi^2$ fit to the cross-section measurements and the vertical dash-dot line corresponds to the date when \ld was at opposition on 2019 June 24 UTC.}
\label{fFiig:lightcurve}
\end{figure}

Using our archival observations of \ld from ZTF taken between 2019 April 9 UTC and 2019 November 8 UTC, we have plotted the equivalent $r$ magnitudes of \ld as a function of time since the perihelion date of 2020 April 10 UTC, ($T - T_p$) measured with equivalent 20,000 km apertures presented in the top panel of Fig.~\ref{fFiig:lightcurve} with observational details in Table~\ref{tab:phot}. These observations include data taken in $g$-band which have been corrected to an equivalent $r$-band magnitude using our $g$-$r$ color estimate for \ld of $\sim$0.6. The 20,000 km aperture was equivalent to an angular size of 5.4\arcsec~ on 2019 November 8 UTC when the comet had a geocentric distance of 5.14 au and an angular size of 7.52\arcsec~ when the comet had a geocentric distance of 3.67 au on 2019 July 1 UTC. The measured local seeing in the ZTF images at the time of observation ranged between 1.6\arcsec~ to 3.9\arcsec~ with a median seeing value of 2.0\arcsec. Using only the r-band photometry, the data show a secular brightening trend of the comet as the comet approached opposition on 2019 June 24.42 UTC increasing in brightness by 1.6$\pm$0.2$\times 10^{-2}$ mags/day. After leaving opposition, the comet showed an asymmetrical secular fading trend of 0.4$\pm$0.1$\times10^{-2}$ mags/day compared to the pre-opposition brightening trend.

In addition to photometry, we present the $\mathrm{A}f\rho$ based off of the ZTF photometry of \ld using as implemented by \citet[][]{Mommert2019ab}. The values $\mathrm{A}f\rho$ are normalized to 0$^{\circ}$ phase angle, $A(0^{\circ})f\rho$ and are plotted against the second Y-axis in the top panel of Fig.~\ref{fFiig:lightcurve} and presented in Table~\ref{tab:phot}. Values of constant $A(0^{\circ})f\rho$ are plotted for reference in the top panel of Fig.~\ref{fFiig:lightcurve}. The value of $A(0^{\circ})f\rho$ rises consistently over the span of our observations resulting in asymmetry in the brightness of \ld as it passed through opposition on 2019 June 24 UTC ($T - T_p$ = -291 days). Before opposition between $T - T_p$ = -370 and -312, the error weighted mean of $A(0^{\circ})f\rho$ = 93.9$\pm$11.8 cm, between $T - T_p$ = -312 and -291, $A(0^{\circ})f\rho$ = 113.1$\pm$8.57 cm. After opposition the error weighted mean of $A(0^{\circ})f\rho$ between $T - T_p$ = -291 and -211 equaled 136.2$\pm$9.7 cm and from $T - T_p$ = -211 and -50, $A(0^{\circ})f\rho$ = 188.8$\pm$25.8 cm. The range of $A(0^{\circ})f\rho$ from 85 cm to 200 cm is consistent with the observed $\mathrm{A}f\rho$ range of comets which ranges from 1 to 10,000 cm \citep[][]{AHearn1995}.

\subsection{Dust properties and mass loss}
\label{sec:dust}

In addition to calculating the $A(0^{\circ})f\rho$ of \ldns, we calculate the value of $C$ for each of the equivalent $r$-band magnitudes in Table~\ref{tab:phot} using Eqn.~\ref{eqn.crosssetion} which are plotted in the bottom panel of Fig.~\ref{fFiig:lightcurve}. A linear function is fit to these data resulting in a fitted slope parameter value of $dC/dt$ = 0.94$\pm$0.06 km$^2$/day. The change in phase angle over the time of our observations is modest as seen in Table~\ref{tab:phot}, therefore variations in the phase function used to calculate $C$ should have a minimal effect on the estimate of the uncertainty of our measured slope parameter. Extrapolating backward in time beyond the range of our data results in a $C$ = 0 km$^2$ at $\sim$135 days before 2019 April 9 UTC, the date of our first photometry data point or on 2018 November 24 UTC during which \ld had a heliocentric distance of $\sim$4.8 au when water ice begins to sublimate \citep[][]{Lisse2019}.

The dimensionless ratio of Solar radiation and gravitational forces is defined by $\beta$ \citep[][]{Burns1979}
\begin{equation}
\beta \, = \, \frac{2L_0r_H^2}{g_{\odot}(1 \mathrm{au})t^2}
\end {equation}
where $L_0$ is the length of dust travel, in this case, the observed length of the tail of \ld of 6.2$\times$10$^{8}$ m, $r_H$ is the heliocentric distance, $g_{\odot}(1 \mathrm{au})$ is the gravitational acceleration towards the Sun at 1 au equal to 6.0$\times$10$^{-3}$ m/s$^2$ and $t$ is the time of particle release. Assuming a mean value of $r_H$$\sim$4.6 au, $L_0$ =  6.2$\times$10$^{8}$ m, the length of the tail estimated from the 2020 April 1 UTC \textit{HST}/WFC3 images, and $t$ = 4.3$\times10^{7}$ s, the time between the 2020 April 1 UTC \textit{HST}/WFC3 observations and the estimated start of the activity, we calculate a value of $\beta$=2.4$\times10^3$. Making the assumption that the dust particles are dielectric spheres \citep[][]{Bohren1983}, the reciprocal of our estimated value of $\beta$ translates into particle size, $\bar{a}$, in $\mu$m of $\sim$400 $\mu$m. However, we caution that this may be an upper limit on the dust size due to the limitations of our tail length measurement by the contamination of background galaxies in the \textit{HST}/WFC3 images due to variations in the activity of \ld affecting our estimate of $t$ based on the backwards extrapolation of the photometric data.

Assuming our estimated particle size of $\sim$400 $\mu$m, we estimate the total mass loss over the duration of our ZTF observations between 2019 April 9 UTC and 2019 November 8 UTC by $M$ = 4/3$\rho\bar{a}\Delta C$ where $\Delta C$ is the difference between the cross section at the start and end of our observations equal to 220 km$^2$. Assuming a dust density of 1 kg/m$^3$ \citep[][]{McDonnell1986}, we obtain a total mass loss over the time span of our observations of $\sim$10$^8$ kg. Adopting our estimated value of $dC/dt$ $\sim$1 km$^2$/day, we obtain a mass loss rate using $\dot{M}$=$4/3\rho\bar{a}$ $dC/dt$$\sim$5$\times$10$^5$kg/day.

To estimate the fraction of active area of \ldns, $A$, we take the ratio between the mass-loss rate and the equilibrium mass sublimation flux at 4.6 au, f$_s$ = 1.4$\times10^{-5}$ kg m$^{-2}$ \citep[][]{Jewitt2015a} where $A$$\sim$0.4 km$^2$. Thus, $\sim$10$\%$ of \ldns's surface is active assuming our inferred size radius of 1.8 km from Section~\ref{sec:morph}, comparable to the active surface area measured for Jupiter family comets \citep[][]{Fernandez1999aap}. An alternative assumption is to assume that \ld has a 100$\%$ active area setting a lower limit to its radius of 0.2 km.

\begin{figure}
\centering
\hspace{-6mm}
\includegraphics[scale=.34]{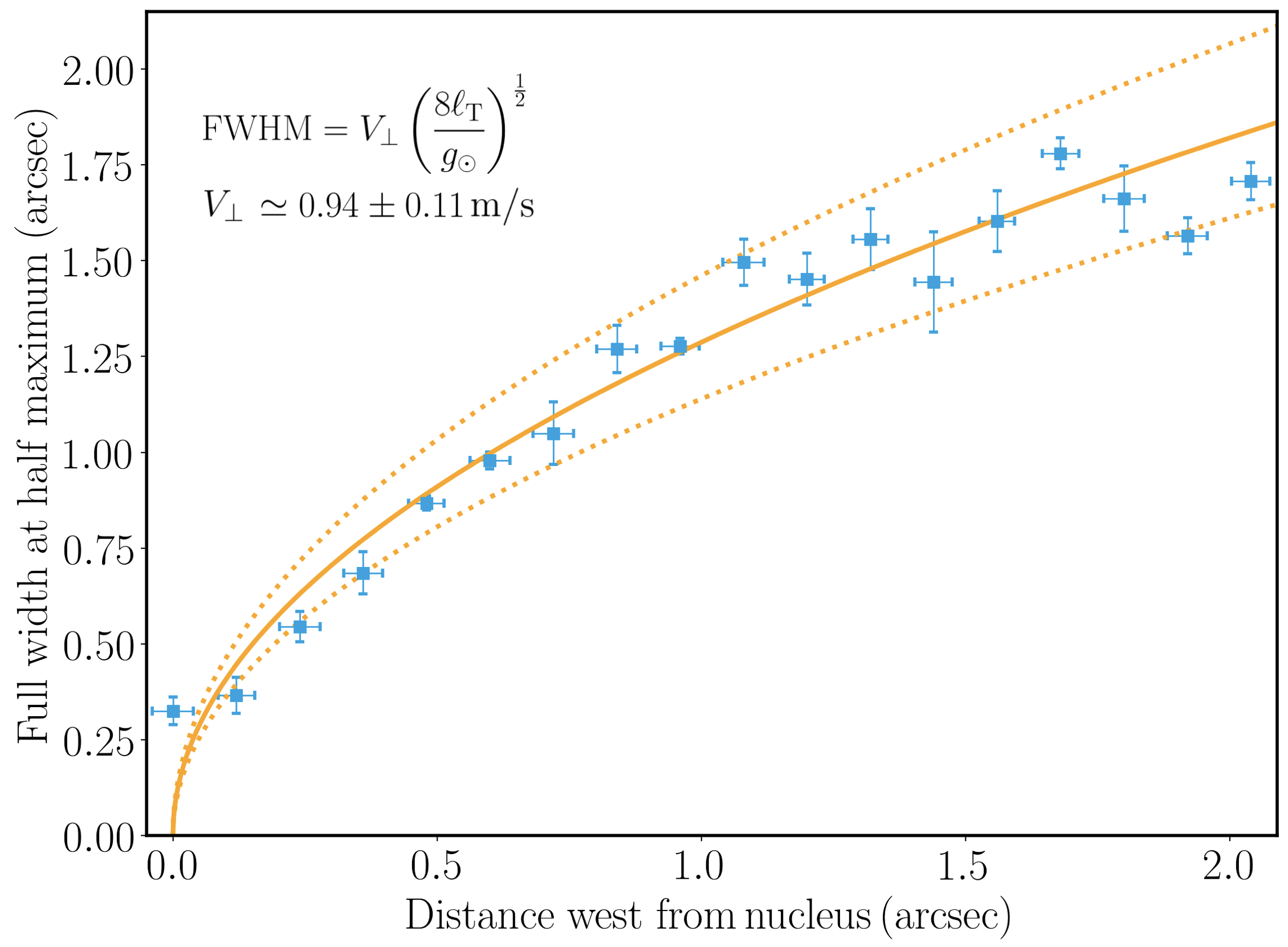}
\caption{The FWHM of the dust tail of \ld vs. the westward angular distance, $\ell_\mathrm{T}$, from the nucleus optocenter of \ld plotted as blue data points when the comet was observed at a -0.44$^\circ$ topo-centric and target orbital plane angle with \textit{HST}/WFC3 on 2020 April 1 UTC. The best fit line in $\ell_\mathrm{T}$ vs. $\mathrm{FWHM}$ space according to Eq.~\ref{eqn.fwhmvsdist} with $V_{\bot}$ = 0.94$\pm$0.11 m/s is plotted as an orange line.}
\label{fFiig:fwhm}
\end{figure}

In addition, we use the perpendicular profile of \ld taken with the high-resolution images from \textit{HST}/WFC3 to estimate the out-of-plane distribution of dust with a minimum of projection effects as the Earth passed through the projected orbital plane of \ld with a projected orbital plane angle of only 0.4$^\circ$ on 2020 April 1 UTC. We measured the FWHM along the perpendicular direction of the tail's profile as a function of distance from the optocenter, $\ell_\mathrm{T}$ between 0 and $\sim$2\arcsec~ in increments of 0.12\arcsec~ slices as are plotted in Fig.~\ref{fFiig:fwhm}.

Neglecting projection effects, the FWHM of the tail gradually widened with $\ell_\mathrm{T}$ and was fit to the following function
\begin{equation}
\label{eqn.fwhmvsdist}
\mathrm{FWHM} = V_{\bot}\left(\frac{8\ell_\mathrm{T}}{g_{\odot}}\right)^{\frac{1}{2}}\end{equation}
from \citep[][]{Jewitt2014} where $V_{\bot}$ is the component of the ejection velocity perpendicular to the orbital plane, equal to  $\sim$1 m/s. We estimate that the perpendicular component of the ejection velocity scales with V$_{\bot}$$\sim$ 1 m/s $(\bar{a}/\bar{a}_0)$$^{-1/2}$ where $\bar{a}_0$ = 400 $\mu$m \citep[][]{Jewitt2014}.

\subsection{Spectrum}
\label{sec:spectrum}

\begin{figure}
\centering
\includegraphics[scale=.425]{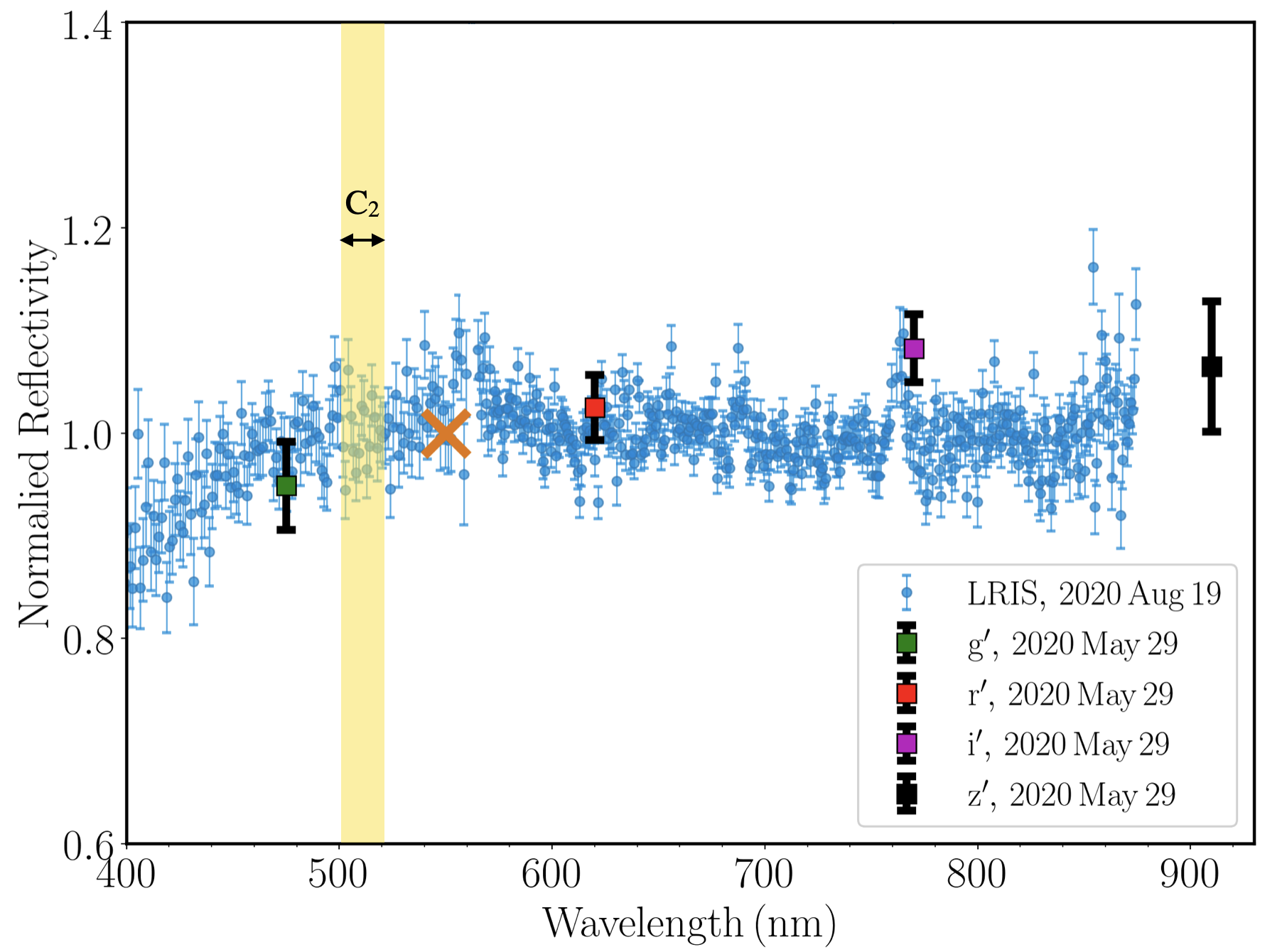}
\caption{Visible wavelength reflectance spectrum taken of \ld with the LRIS instrument on Keck I on 2020 August 19 plotted as blue dots. The error bars on the spectrum data points correspond to 1-$\sigma$ uncertainty. The spectrum has been normalized to unity at 550 nm indicated by the orange cross. The spectrum presented was obtained by combining two spectra from the blue camera using the 600/4000 grism and the red camera using the 600/7500 grating with a 560 nm dichroic \cite{Oke1995,McCarthy1998}. The data have been rebinned and smoothed by a factor of 10 using an error-weighted mean. The spectral range of the cometary C$_2$ emission line has been indicated by the yellow shaded area \citep[][]{Farnham2000}. The spike in the spectrum at $\sim$560 nm is due an artifact caused by the dichroic solution and the spike at $\sim$760 nm is caused by the telluric H$_2$O absorption feature in both the comet and Solar analog spectrum. The data to reproduce our plot of the reflectivity spectrum of \ld are available at the following link: \href{https://sites.astro.caltech.edu/~bbolin/Keck_I_LRIS_P_2019_LD2_2020_Aug_19_UTC_wavelength_nm_reflectivity_reflectivity_1_sigma_uncertainty.txt}{[link]}.}
\label{fFiig:spectrum}
\end{figure}

The spectrum of \ld was extracted using a 7.4\arcsec-wide region centered on the peak of the continuum's brightness. We compute the normalized reflectance spectrum of \ld taken with Keck/LRIS on 2020 August 19 UTC in the wavelength range between 400 nm and 1000 nm by dividing the \ld spectrum by our Solar analog spectrum and normalized to unity at 550 nm. The resulting spectrum indicates a reddish to neutral coma color for \ld as seen in Fig.~\ref{fFiig:spectrum} . We measured a slope of $\sim$16$\%$/100 nm between 480 and 760 nm and a flatter spectrum between 760 nm and 900 nm consistent with the photometric colors taken by LT on 2020 May 29 plotted over the LRIS spectra in Fig.~\ref{fFiig:spectrum} for reference. Our spectrum shows no sign of $C_2$ emission in the 505 nm to 522 nm range \citep[][]{Farnham2000} in the highlighted range in Fig.~\ref{fFiig:spectrum}.

We set an upper limit to the $C_2$ gas production of \ld using the mean $V$-band continuum flux density of \ld using its measured 550 nm flux, flux$_V$ = 1.52$\times10^{-15}$ erg cm$^{-2}$ s$^{-1}$ nm$^-1$. The fractional 1$\sigma$ continuum statistical uncertainty of our \ld spectrum in the range spanning the 505 nm to 522 nm is 0.01 corresponding to a 3$\sigma$ flux density flux$_\mathrm{C_2}$ = 2.13$\times10^{-17}$ erg cm$^{-2}$ s$^{-1}$ nm$^-1$ including a correction of 0.6 for slit losses. The 3$\sigma$ upper limit to the flux in the 17 nm width of the C$_2$ band is $F_\mathrm{C_2}\leq$ 3.62$\times10^{-16}$ erg cm$^{-2}$ s$^{-1}$. The 3$\sigma$ upper limit on the number of $C_2$ molecules projected within the 7.4\arcsec~ $\times$ 1.0\arcsec~spectroscopic slit assuming that the coma is optically thin is
\begin{equation}
N_{mol} = \frac{4\pi\Delta^2F_\mathrm{C_2}}{g(r_h)}
\end{equation}
where $g(r_h)$ is the fluorescence efficiency factor of the $C_2$ spectral band at $r_h$ where $g(1\,\mathrm{au})$ = 2.2$\times10^{-14}$ erg s$^{-1}$ radical$^{-1}$ \citep[][]{AHearn1982}
 which results in $N_{mol}\leq$1.27$\times10^{27}$ molecules.
 
We apply the assumptions of the Haser model \citep[][]{Haser1957} to determine a coarse 3$\sigma$ upper limit on the production rate of $C_2$. The Haser model uses two length scales, the "parent" molecule species length scale, $L_P$, and the "daughter" molecule species length scale, $L_D$, to describe the distribution of the radicals. For $C_2$ at a $r_h$ of 4.594 au, $L_P$ =5.3$\times10^{5}$ km and $L_D$ =2.5$\times10^{6}$ km \citep[][]{Cochran1985}. In addition, we assume the speed of the molecular gas is 0.5 km/s which is used to determine the residence time of the molecules in the projected slit \citep[][]{Combi2004}. Using these assumptions with the Haser model we find the 3$\sigma$ upper limit to the gas production rate $Q_{C_2}\lesssim$7.5$\times10^{24}$ mol/s which is a similar limit compared to the measured $Q_{C_2}$ of other Solar System Comets at similar heliocentric distances \citep[][]{Feldman2004} and to the results of \citet[][]{Licandro2020DPS}. Scaling our measured spectroscopic upper limit on the $Q_{C_2}$ gas production rate to a OH gas production rate using the median ratio of $C_2$ to hydroxyl production rate for Solar System comets \citep[][]{AHearn1995} results in an estimated spectroscopic upper limit to $Q_{OH}$$\lesssim$2.4$\times10^{27}$ mol/s and an mass loss rate in water of $dM_{H_2O}/dt$$\lesssim$80 kg/s.

\subsection{Orbital evolution}

In order to investigate the long-term dynamics of 2019 LD$_2$, we simulated 27,000 clones of its orbit. The clone set is created by using 1,000 three-dimentional locations using the positional uncertainties. The velocity uncertainties are accounted for by creating 27 clones in the three-dimensional velocity space at each positional location for a total of 27,000 clones. Orbital six-vectors were generated with uncertainties from the JPL Horizons Orbit Solution dated 2020 May 20 00:43:28 and set for the 2019 September 15 00:00 UTC epoch. In addition, we use the major gravitational components of the Solar system (Sun, Venus, Earth, Mars, Jupiter, Saturn, Uranus and Neptune.). The simulations were conducted using {\tt REBOUND} \citep[][]{Rein2012REBOUND} with the hybrid {\tt MERCURIUS} integrator \citep[][]{Rein2019MERCURIUS}. We ran two sets of integrations, a short-term set, integrated for 100 years, and a long-term set for $1 \times 10^{7}$ years. For each simulation set, we use a time-step size of 0.025 years. For the long-term integrations, we output every 1,000 years and analyze the time at which the clones escape the Solar System (distance from Sun larger than 1000~au).

\begin{figure}
\centering
\includegraphics[scale=.55]{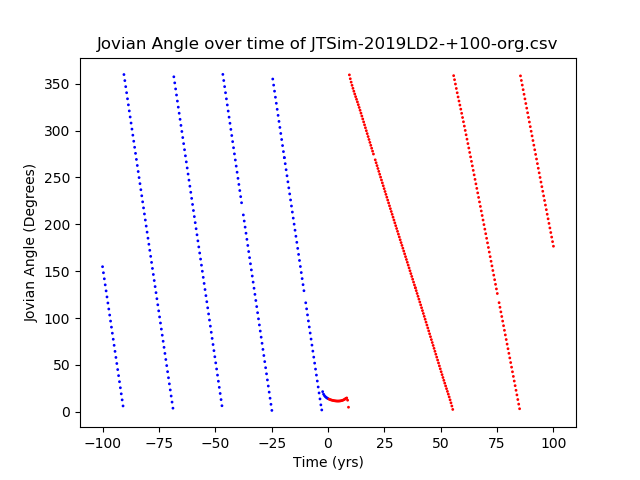}
\includegraphics[scale=.55]{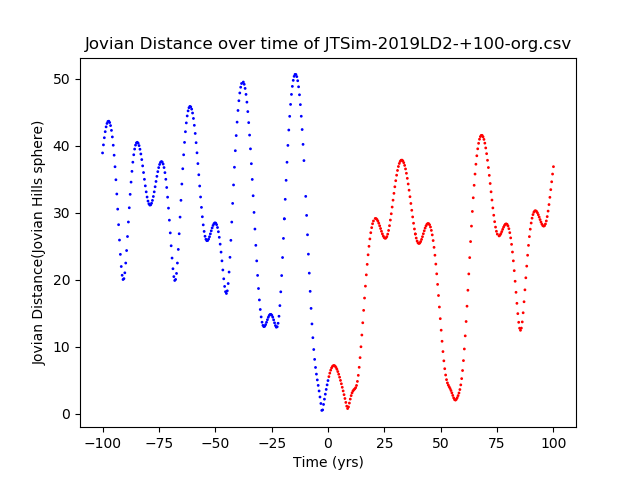}
\includegraphics[scale=.55]{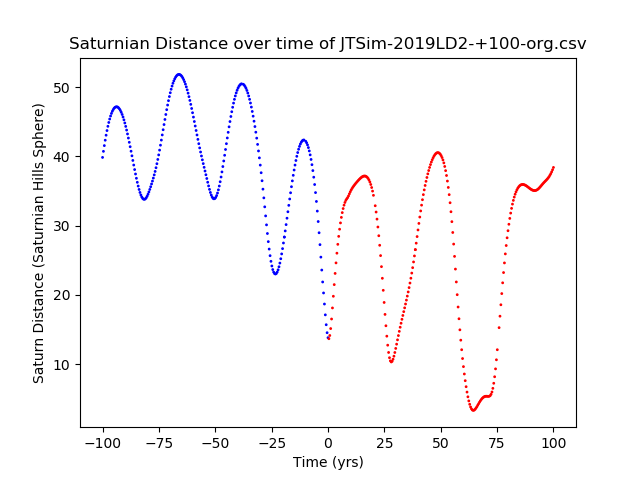}
\includegraphics[scale=.55]{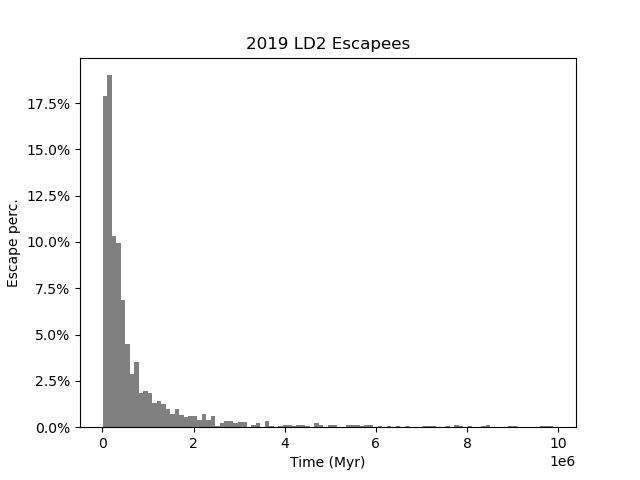}

\caption{Top left panel: Jovian angle defined as the relative longitude between Jupiter and \ld as a function of time between -100 and 100 years centered on 2019 September 15 UTC. Except for a brief period of time between -3 and 8 years where the Jovian angle was $\sim$20$^\circ$, the Jovian angle cycled between 0$^\circ$ and 360$^\circ$. Top right panel:  same as the top left panel except for the Jovian distance defined as the distance between \ld and Jupiter in Jovian hill spheres ($\sim$0.35 au) as a function of time. The local minimum in the Jovian distance occurred $\sim$2.77 years ago with a distance of $\sim$0.50 Jovian hill radius, or 0.17 au. Bottom left panel: same as the top right panel except for the Saturn distance defined as the distance between \ld and Saturn in Saturn Hill spheres (0.43 au). The local minimum occurs in $\sim$60 years when \ld comes within $\sim$3.3 Hill radii or 1.42 au of Saturn. Bottom right panel: the percentage of orbital \ld clones that have escaped the Solar System (reached $>$1000 au from the Sun) per bin in duration of time. Each bin is $\sim$100,000 years wide. Within the first million years of the simulation, $\sim$78.8$\%$ have escaped the Solar System. By 10 million years, $\sim$95$\%$ have escaped the Solar System.}
\label{fFiig:orbit}
\end{figure}

The short-term integrations replicate the previous work of \citet[][]{Steckloff2020} and \citet[][]{Hsieh20202019ld2}. In the simulations, we find that the clones entered the Jovian region approximately 2.37 years ago, and are ejected from the region 8.70 years in the future (See top left panel of Fig.~\ref{fFiig:orbit} for a clone example). As was found previously, 2019 LD$_2$ transitions from a Centaur, with an approximate semi-major axis of 8.6~au to a Jupiter family comet with a semi-major axis of 6.2~au, spending 11.0712 years in the Jovian region (semi-major axis 5.2~au). In the long-term simulations ($1 \times 10^{7}$ years) we find that half of the clones escape the Solar System in 3.4 $\times 10^5$ years and 78.8$\%$ of the clones escape the Solar system within the first $1 \times 10^{6}$ years as seen in the bottom-right panel of Fig.~\ref{fFiig:orbit}. After $\sim$3.8 Myrs, 95$\%$ of the \ld clones have escaped.

\section{Discussion and Conclusions}
From our observations spanning multiple observatories, transitioning comet \ld exhibits interesting features in comparison with other short period Solar System comets. \ld has a higher value of $\mathrm{A}f\rho$ of 85-200 cm \citep[][]{AHearn1995} at a helio-centric distance between 4.7 and 4.6 au compared to other short period comets at a similar helio-centric distance \citep[][]{Kelley2013,Ivanova2014,Bauer2015} and a $\sim$km-scale nucleus \citep[][]{Fernandez2013} as well as reddish to neutral color properties \citep[][]{Jewitt2015bbb}. The comet's morphology when observed over multiple epochs since 2019 April exhibits the presence of a tail suggesting sustained activity versus an impulsive event. In addition, \ld has a moderate upper limit for the production of CO/CO$_2$ of $\sim$10$^{27}$ mol/s and $\sim$10$^{26}$ mol/s, respectively, based on our \textit{Spitzer} observations taken in 2020 January. It is also very active when compared to 29P (although 29P is located at a slightly farther heliocentric distance of $\sim$6 au) on a per unit surface area basis ($\mathrm{A}f\rho$$\sim$150cm/(1.8 km)$^2$$\sim$50 cm/km$^2$ for \ldns) vs \citep[$\mathrm{A}f\rho$$\sim$1000 cm /(30.5 km)$^2$$\sim$1 cm/km$^2$ for for 29P, ][]{Ivanova2011}, using the latest value for SW1's size from \citet[][]{Schambeau2019}. But this activity seems to produce quite large ($\sim$100 $\mu$m) reddish dust particles containing copious amounts of water ice according to our work and that of \citet[][]{Kareta2020ld2}.

In addition to the morphology of the comet indicating sustained activity, photometry of the comet observed between 2019 April and 2019 November by ZTF is consistent with the activity steadily increasing since late 2018 up through the end of 2019 and into 2020 as the comet nears its perihelion on 2020 April 10 UTC. The length of the tail in deep \text{HST} imaging as well as our inferred start of activity date of late 2018 implies that the coma consists of $\sim$100 $\mu$m-scale dust ejected at a relatively low velocity of $\sim$1 m/s. Although this is roughly consistent with the escape speed of a non-rotating comet nucleus of radius $\sim$1.8 km as inferred by our observations, it is unlikely that the dust is being ejected exclusively by rotational mass shedding suggested by the low ejection velocity \citep[e.g.][]{Ye2019adfs,Lin2020gault}. Rather, the increased activity of the comet as it nears perihelion suggest that dust is being transported by the sublimation, or that the activity is a product of both the sublimation of volatiles and rotational mass shedding.

The size of the active region on \ld of $\sim$0.4 km$^2$ is too large to explain the low ejection velocity of the dust as for other comets with low dust ejection velocities whose activity is driven by the sublimation of volatiles \citep[][]{Jewitt2014}. Subsequent observations of \ld to determine the rotation state of the comet will be necessary to understand if it is rotating near its critical rotation limit indicating the role of rotational mass shedding in its activity or if the comet has the possibility of becoming rotationally disrupted in the near future or if it has disrupted in the recent past \citep[][]{Moreno2017,Vokrouhlicky2017a}. Given the 4-5 au location of the comet during its recent epoch of activity, the distance at which water ice begins to sublimate \citep[][]{Meech2004}, it is likely that the activity is being driven primarily by the sublimation of water ice. An additional possible mechanism may be the transformation of amorphous water ice into crystalline water as has been suggested before as an activity-driving mechanism for Centaurs \citep[][]{Jewitt2009}. It is possible for other volatiles such as CO to partially drive the activity of \ld as seen in other distant comets \citep[e.g.,][]{Bolin2020asdfasdf} and 29P \citep[][]{Gunnarsson2008ada}, however the lack of detection of the activity at large heliocentric distances \citep[][]{Schambeau2020cbet} seems to suggest that hyper-volatiles are not the dominant drivers of the activity of \ldns. If the lack of hyper-volatiles driving the activity of \ld is confirmed, it may suggest that \ld has spent a significant amount of time as a Centaur within 15 au of the Sun \citep[][]{Horner2004CentaursI} where these hyper-volatiles may have had a greater chance to become depleted compared to water ice which is non-volatile at that distance. 

Additionally, some comets show evidence of a strong transition between H$_2$O-driven and CO-driven activity at heliocentric distances past $\sim$3.5 au such as for comets 67P \citep[][]{Lauter2019} and Hale-Bopp \citep[][]{Biver1997}. Our observation of the activity of \ld and its inferred H$_2$O-driven activity at its heliocentric distance of $\sim$4.6 au at the time of our observations is seemingly at odds with the transition to CO-driven activity at larger heliocentric distances as observed for other comets. However, it has been shown that the shape and rotation pole orientation of comets can have a strong effect on the distance at which comet activity is driven by H$_2$O. H$_2$O-driven activity can increase at larger heliocentric distances for comet shapes and orientations deviating from a spherically-shaped comet rotating perpendicular to its orbital plane \citep[][]{Marshall2019}. Therefore, the inferred H$_2$O-driven activity of \ld at its large heliocentric distance of 4.6 au could be explained by it having a non-spherical shape and significant obliquity which has been shown to sustain H$_2$O activity at these heliocentric distances in comet-activity models.

\ld is beginning to enter the region where water ice begins to appreciably sublimate at rates high enough so that a patch of pure water ice in the surface would disappear on year-long timescales \citep[][]{Lisse2020}. The beginnings of mobilization of water ice for a weakly structured surface such as found on 67P \citep[][]{ORourke2020} could lead to the slow flaking off of large chunks of the loosest, weakest material that had never felt such stresses before. In this scenario gas will evolve at low levels in order to drive dust off the object. The material driven off should be water ice rich, as this ice is the last, most refractory ice expected in a cometary body before it is totally volatile ice depleted. We then may expect to see \ldns's activity modulated by its motion towards/away from the Sun over an orbit similar to how the activity of Main Belt Comets are modulated as they travel inside/outside the 2.5 au water ``ice line'' where water ice boils furiously into vacuum \citep[][]{Hsieh2015,Hsieh2015a}, \ldns's activity could be modulated by its traversing the water ice turn on line of activity. Future monitoring observations over the next years will determine if this is the case, as is suggested by the smoothly increasing $\mathrm{A}f\rho$ towards perihelion values we find for \ldns, they do not appear to be describing an impulsive outburst.

In our long-term simulations ($1 \times 10^{7}$ years), we show the temporary nature of 2019 LD$_2$. Up to 78.8$\%$ of our orbital clones escape in the first $1 \times 10^{6}$ years. The half-life of the clones is approximately 3.4 $\times 10^5$ years. This is an order of magnitude smaller than the mean half-life of Centaurs \citep[$2.7 \times 10^{6}$ years ][]{Horner2004CentaursI} and more comparable to lifetimes of Jupiter Family group comets of $\sim 5 \times 10^{5}$ years \citep[][]{Levison1994}.

Without a robust assessment of survey selection effects \citep[e.g.,][]{Jedicke2016,Boe2019} it is difficult to assess the true population of comets in a temporary co-orbital configuration with Jupiter and transitioning between the Centaur and Jupiter family comet populations \citep[][]{Sarid2019}. However, we can use our estimated size of \ld in comparison with the population estimates of Centaurs in the transition region \citep[][]{Steckloff2020}. \citet[][]{Steckloff2020} predict that there are $\sim$40-1,000 objects in the transition region with radius $>$1-3 km with fewer if cometary fading is considered in the population estimate \citep[][]{Brasser2015}. Using these transition object population estimates and our estimate of the radius of \ld of $\sim$1.8 km suggests there are $\sim$100 objects the size of \ld in the transition region at any given time. Additional monitoring of \ld and objects like it in the gateway region will be required to understand their activity drivers and population.

\acknowledgments

The authors wish to thank the two anonymous reviewers for their help in revising this manuscript which greatly improved its text.

Based on observations with the NASA/ESA Hubble Space Telescope obtained from the Data Archive at the Space Telescope Science Institute, which is operated by the Association of Universities for Research in Astronomy, Incorporated, under NASA contracts NAS5-26555. Support for Program number (GO 16077) was provided through a grant from the STScI under NASA contract NAS5-26555.

This work is based on observations made with the Spitzer Space Telescope, which was operated by the Jet Propulsion Laboratory, California Institute of Technology, under a contract with NASA.

Based on observations obtained with the Samuel Oschin Telescope 48-inch and the 60-inch Telescope at the Palomar Observatory as part of the Zwicky Transient Facility project. ZTF is supported by the National Science Foundation under Grant No. AST-1440341 and a collaboration including Caltech, IPAC, the Weizmann Institute for Science, the Oskar Klein Center at Stockholm University, the University of Maryland, the University of Washington, Deutsches Elektronen-Synchrotron and Humboldt University, Los Alamos National Laboratories, the TANGO Consortium of Taiwan, the University of Wisconsin at Milwaukee, and Lawrence Berkeley National Laboratories. Operations are conducted by COO, IPAC, and UW.

The Liverpool Telescope is operated on the island of La Palma by Liverpool John Moores University in the Spanish Observatorio del Roque de los Muchachos of the Instituto de Astrofisica de Canarias with financial support from the UK Science and Technology Facilities Council.

This work was supported by the GROWTH project funded by the National Science Foundation under PIRE Grant No 1545949.

Some of the data presented herein were obtained at the W. M. Keck Observatory, which is operated as a scientific partnership among the California Institute of Technology, the University of California and the National Aeronautics and Space Administration. The Observatory was made possible by the generous financial support of the W. M. Keck Foundation.

The authors wish to recognize and acknowledge the very significant cultural role and reverence that the summit of Maunakea has always had within the indigenous Hawaiian community. We are most fortunate to have the opportunity to conduct observations from this mountain.

B.T.B., G.H. and F.J.M. acknowledge support from NASA with grant number 80NSSC19K0780.

C.F.~gratefully acknowledges the support of his research by the Heising-Simons Foundation ($\#$2018-0907).

M.~W.~Coughlin acknowledges support from the National Science Foundation with grant number PHY-2010970.

This publication has made use of data collected at Lulin Observatory, partly supported by MoST grant 108-2112-M-008-001.

C.C.N. thanks to the funding from MOST grant 104-2923-M-008-004-MY5.

The authors would like to acknowledge the helpful discussion of \ld with L. Woodney and Q.-Z. Ye.

This work has made use of data from the European Space Agency (ESA) mission \textit{Gaia} (\texttt{https://www.cosmos.esa.int/gaia}), processed by the \textit{Gaia} Data Processing and Analysis Consortium (DPAC, \texttt{https://www.cosmos.esa.int/web/gaia/dpac/\\consortium}). Funding for the DPAC has been provided by national institutions, in particular the institutions participating in the \textit{Gaia} Multilateral Agreement.

\facility{\textit{Hubble Space Telescope}, \textit{Spitzer Space Telescope}, Keck I Telescope, P48 Oschin Schmidt telescope/Zwicky Transient Facility, Apache Point Astrophysical Research Consortium 3.5 m telescope, Liverpool Telescope, Lulin Optical Telescope, Mount Laguna Observatory 40-inch Telescope}
\software{Small Body Python, ZChecker, LPipe, REBOUND}

\bibliographystyle{aasjournal}
\bibliography{ms}

\begin{thebibliography}{}
\expandafter\ifx\csname natexlab\endcsname\relax\def\natexlab#1{#1}\fi
\providecommand{\url}[1]{\href{#1}{#1}}
\providecommand{\dodoi}[1]{doi:~\href{http://doi.org/#1}{\nolinkurl{#1}}}
\providecommand{\doeprint}[1]{\href{http://ascl.net/#1}{\nolinkurl{http://ascl.net/#1}}}
\providecommand{\doarXiv}[1]{\href{https://arxiv.org/abs/#1}{\nolinkurl{https://arxiv.org/abs/#1}}}

\bibitem[{{A'Hearn}(1982)}]{AHearn1982}
{A'Hearn}, M.~F. 1982, in IAU Colloq. 61: Comet Discoveries, Statistics, and
  Observational Selection, ed. L.~L. {Wilkening}, 433--460

\bibitem[{{A'Hearn} {et~al.}(1995){A'Hearn}, {Millis}, {Schleicher}, {Osip}, \&
  {Birch}}]{AHearn1995}
{A'Hearn}, M.~F., {Millis}, R.~C., {Schleicher}, D.~O., {Osip}, D.~J., \&
  {Birch}, P.~V. 1995, \icarus, 118, 223, \dodoi{10.1006/icar.1995.1190}

\bibitem[{{A'Hearn} {et~al.}(1984){A'Hearn}, {Schleicher}, {Millis}, {Feldman},
  \& {Thompson}}]{AHearn1984}
{A'Hearn}, M.~F., {Schleicher}, D.~G., {Millis}, R.~L., {Feldman}, P.~D., \&
  {Thompson}, D.~T. 1984, \aj, 89, 579, \dodoi{10.1086/113552}

\bibitem[{{Bauer} {et~al.}(2013){Bauer}, {Grav}, {Blauvelt}, {Mainzer},
  {Masiero}, {Stevenson}, {Kramer}, {Fern{\'a}ndez}, {Lisse}, {Cutri},
  {Weissman}, {Dailey}, {Masci}, {Walker}, {Waszczak}, {Nugent}, {Meech},
  {Lucas}, {Pearman}, {Wilkins}, {Watkins}, {Kulkarni}, {Wright}, {WISE Team},
  \& {PTF Team}}]{Bauer2013}
{Bauer}, J.~M., {Grav}, T., {Blauvelt}, E., {et~al.} 2013, \apj, 773, 22,
  \dodoi{10.1088/0004-637X/773/1/22}

\bibitem[{{Bauer} {et~al.}(2015){Bauer}, {Stevenson}, {Kramer}, {Mainzer},
  {Grav}, {Masiero}, {Fern{\'a}ndez}, {Cutri}, {Dailey}, {Masci}, {Meech},
  {Walker}, {Lisse}, {Weissman}, {Nugent}, {Sonnett}, {Blair}, {Lucas},
  {McMillan}, {Wright}, {WISE}, \& {NEOWISE Teams}}]{Bauer2015}
{Bauer}, J.~M., {Stevenson}, R., {Kramer}, E., {et~al.} 2015, \apj, 814, 85,
  \dodoi{10.1088/0004-637X/814/2/85}

\bibitem[{{Bellm} {et~al.}(2019){Bellm}, {Kulkarni}, {Graham}, {Dekany},
  {Smith}, {Riddle}, {Masci}, {Helou}, {Prince}, {Adams}, {Barbarino},
  {Barlow}, {Bauer}, {Beck}, {Belicki}, {Biswas}, {Blagorodnova}, {Bodewits},
  {Bolin}, {Brinnel}, {Brooke}, {Bue}, {Bulla}, {Burruss}, {Cenko}, {Chang},
  {Connolly}, {Coughlin}, {Cromer}, {Cunningham}, {De}, {Delacroix}, {Desai},
  {Duev}, {Eadie}, {Farnham}, {Feeney}, {Feindt}, {Flynn}, {Franckowiak},
  {Frederick}, {Fremling}, {Gal-Yam}, {Gezari}, {Giomi}, {Goldstein},
  {Golkhou}, {Goobar}, {Groom}, {Hacopians}, {Hale}, {Henning}, {Ho}, {Hover},
  {Howell}, {Hung}, {Huppenkothen}, {Imel}, {Ip}, {Ivezi{\'c}}, {Jackson},
  {Jones}, {Juric}, {Kasliwal}, {Kaspi}, {Kaye}, {Kelley}, {Kowalski},
  {Kramer}, {Kupfer}, {Landry}, {Laher}, {Lee}, {Lin}, {Lin}, {Lunnan},
  {Giomi}, {Mahabal}, {Mao}, {Miller}, {Monkewitz}, {Murphy}, {Ngeow},
  {Nordin}, {Nugent}, {Ofek}, {Patterson}, {Penprase}, {Porter}, {Rauch},
  {Rebbapragada}, {Reiley}, {Rigault}, {Rodriguez}, {van Roestel}, {Rusholme},
  {van Santen}, {Schulze}, {Shupe}, {Singer}, {Soumagnac}, {Stein}, {Surace},
  {Sollerman}, {Szkody}, {Taddia}, {Terek}, {Van Sistine}, {van Velzen},
  {Vestrand}, {Walters}, {Ward}, {Ye}, {Yu}, {Yan}, \& {Zolkower}}]{Bellm2019}
{Bellm}, E.~C., {Kulkarni}, S.~R., {Graham}, M.~J., {et~al.} 2019, 131, 018002,
  \dodoi{10.1088/1538-3873/aaecbe}

\bibitem[{{Biver} {et~al.}(1997){Biver}, {Bockel{\'e}e-Morvan}, {Colom},
  {Crovisier}, {Germain}, {Lellouch}, {Davies}, {Dent}, {Moreno}, {Paubert},
  {Wink}, {Despois}, {Lis}, {Mehringer}, {Benford}, {Gardner}, {Phillips},
  {Gunnarsson}, {Rickman}, {Winnberg}, {Bergman}, {Johansson}, \&
  {Rauer}}]{Biver1997}
{Biver}, N., {Bockel{\'e}e-Morvan}, D., {Colom}, P., {et~al.} 1997, Earth Moon
  and Planets, 78, 5, \dodoi{10.1023/A:1006229818484}

\bibitem[{{Boe} {et~al.}(2019){Boe}, {Jedicke}, {Meech}, {Wiegert}, {Weryk},
  {Chambers}, {Denneau}, {Kaiser}, {Kudritzki}, {Magnier}, {Wainscoat}, \&
  {Waters}}]{Boe2019}
{Boe}, B., {Jedicke}, R., {Meech}, K.~J., {et~al.} 2019, \icarus, 333, 252,
  \dodoi{10.1016/j.icarus.2019.05.034}

\bibitem[{{Bohren} \& {Huffman}(1983)}]{Bohren1983}
{Bohren}, C.~F., \& {Huffman}, D.~R. 1983, {Absorption and scattering of light
  by small particles}

\bibitem[{{Bolin} {et~al.}(2019){Bolin}, {Fernandez}, {Bauer}, {Helou}, \&
  {Lisse}}]{Bolin2019Spitzer}
{Bolin}, B., {Fernandez}, Y., {Bauer}, J., {Helou}, G., \& {Lisse}, C. 2019,
  {Spitzer investigation of the first known active Trojan, 2019 LD2}, Spitzer
  Proposal

\bibitem[{{Bolin} {et~al.}(2020{\natexlab{a}}){Bolin}, {Bodewits}, {Fernandez},
  \& {Lisse}}]{BolinHST2020}
{Bolin}, B.~T., {Bodewits}, D., {Fernandez}, Y., \& {Lisse}, C.~M.
  2020{\natexlab{a}}, {Determining the cause of activity of the first active
  Trojan, 2019 LD2}, HST Proposal

\bibitem[{{Bolin} \& {Lisse}(2020)}]{Bolin2020HST}
{Bolin}, B.~T., \& {Lisse}, C.~M. 2020, \mnras, 497, 4031,
  \dodoi{10.1093/mnras/staa2192}

\bibitem[{{Bolin} {et~al.}(2018){Bolin}, {Weaver}, {Fernandez}, {Lisse},
  {Huppenkothen}, {Jones}, {Juri{\'c}}, {Moeyens}, {Schambeau}, {Slater},
  {Ivezi{\'c}}, \& {Connolly}}]{Bolin2018}
{Bolin}, B.~T., {Weaver}, H.~A., {Fernandez}, Y.~R., {et~al.} 2018, \apjl, 852,
  L2, \dodoi{10.3847/2041-8213/aaa0c9}

\bibitem[{{Bolin} {et~al.}(2020{\natexlab{b}}){Bolin}, {Lisse}, {Kasliwal},
  {Quimby}, {Tan}, {Copperwheat}, {Lin}, {Morbidelli}, {Abe}, {Bendjoya},
  {Burdge}, {Coughlin}, {Fremling}, {Itoh}, {Koss}, {Masci}, {Maeno},
  {Mamajek}, {Marocco}, {Murata}, {Rivet}, {Sitko}, {Stern}, {Vernet},
  {Walters}, {Yan}, {Andreoni}, {Bhalerao}, {Bodewits}, {De}, {Deshmukh},
  {Bellm}, {Blagorodnova}, {Buzasi}, {Cenko}, {Chang}, {Chojnowski}, {Dekany},
  {Duev}, {Graham}, {Juri{\'c}}, {Kulkarni}, {Kupfer}, {Mahabal}, {Neill},
  {Ngeow}, {Penprase}, {Riddle}, {Rodriguez}, {Smith}, {Rosnet}, {Sollerman},
  \& {Soumagnac}}]{Bolin2020asdfasdf}
{Bolin}, B.~T., {Lisse}, C.~M., {Kasliwal}, M.~M., {et~al.} 2020{\natexlab{b}},
  \aj, 160, 26, \dodoi{10.3847/1538-3881/ab9305}

\bibitem[{{Brasser} \& {Wang}(2015)}]{Brasser2015}
{Brasser}, R., \& {Wang}, J.~H. 2015, \aap, 573, A102,
  \dodoi{10.1051/0004-6361/201423687}

\bibitem[{{Burns} {et~al.}(1979){Burns}, {Lamy}, \& {Soter}}]{Burns1979}
{Burns}, J.~A., {Lamy}, P.~L., \& {Soter}, S. 1979, \icarus, 40, 1,
  \dodoi{10.1016/0019-1035(79)90050-2}

\bibitem[{{Chandler} {et~al.}(2020){Chandler}, {Kueny}, {Trujillo}, {Trilling},
  \& {Oldroyd}}]{Chandler2020}
{Chandler}, C.~O., {Kueny}, J.~K., {Trujillo}, C.~A., {Trilling}, D.~E., \&
  {Oldroyd}, W.~J. 2020, \apjl, 892, L38, \dodoi{10.3847/2041-8213/ab7dc6}

\bibitem[{{Cochran}(1985)}]{Cochran1985}
{Cochran}, A.~L. 1985, \aj, 90, 2609, \dodoi{10.1086/113966}

\bibitem[{{Combi} {et~al.}(2004){Combi}, {Harris}, \& {Smyth}}]{Combi2004}
{Combi}, M.~R., {Harris}, W.~M., \& {Smyth}, W.~H. 2004, {Gas dynamics and
  kinetics in the cometary coma: theory and observations}, ed. M.~C. {Festou},
  H.~U. {Keller}, \& H.~A. {Weaver}, 523

\bibitem[{{De Sanctis} {et~al.}(2000){De Sanctis}, {Capria}, {Coradini}, \&
  {Orosei}}]{DeSanctis2000}
{De Sanctis}, M.~C., {Capria}, M.~T., {Coradini}, A., \& {Orosei}, R. 2000,
  \aj, 120, 1571, \dodoi{10.1086/301512}

\bibitem[{{DeMeo} \& {Carry}(2013)}]{DeMeo2013}
{DeMeo}, F.~E., \& {Carry}, B. 2013, \icarus, 226, 723,
  \dodoi{10.1016/j.icarus.2013.06.027}

\bibitem[{{Deustua} {et~al.}(2017){Deustua}, {Mack}, {Bajaj}, \&
  {Khandrika}}]{Deustua2017}
{Deustua}, S.~E., {Mack}, J., {Bajaj}, V., \& {Khandrika}, H. 2017, {WFC3/UVIS
  Updated 2017 Chip-Dependent Inverse Sensitivity Values}, Tech. rep.

\bibitem[{{Dones} {et~al.}(2015){Dones}, {Brasser}, {Kaib}, \&
  {Rickman}}]{Dones2015}
{Dones}, L., {Brasser}, R., {Kaib}, N., \& {Rickman}, H. 2015, \ssr, 197, 191,
  \dodoi{10.1007/s11214-015-0223-2}

\bibitem[{{Dressel}(2012)}]{Dressel2012}
{Dressel}, L. 2012, {Wide Field Camera 3 Instrument Handbook for Cycle 21 v.
  5.0}

\bibitem[{{Duncan} {et~al.}(2004){Duncan}, {Levison}, \& {Dones}}]{Duncan2004}
{Duncan}, M., {Levison}, H., \& {Dones}, L. 2004, {Dynamical evolution of
  ecliptic comets}, ed. M.~C. {Festou}, H.~U. {Keller}, \& H.~A. {Weaver}, 193

\bibitem[{{Farnham} {et~al.}(2000){Farnham}, {Schleicher}, \&
  {A'Hearn}}]{Farnham2000}
{Farnham}, T.~L., {Schleicher}, D.~G., \& {A'Hearn}, M.~F. 2000, \icarus, 147,
  180, \dodoi{10.1006/icar.2000.6420}

\bibitem[{{Fazio} {et~al.}(2004){Fazio}, {Hora}, {Allen}, {Ashby}, {Barmby},
  {Deutsch}, {Huang}, {Kleiner}, {Marengo}, {Megeath}, {Melnick}, {Pahre},
  {Patten}, {Polizotti}, {Smith}, {Taylor}, {Wang}, {Willner}, {Hoffmann},
  {Pipher}, {Forrest}, {McMurty}, {McCreight}, {McKelvey}, {McMurray}, {Koch},
  {Moseley}, {Arendt}, {Mentzell}, {Marx}, {Losch}, {Mayman}, {Eichhorn},
  {Krebs}, {Jhabvala}, {Gezari}, {Fixsen}, {Flores}, {Shakoorzadeh}, {Jungo},
  {Hakun}, {Workman}, {Karpati}, {Kichak}, {Whitley}, {Mann}, {Tollestrup},
  {Eisenhardt}, {Stern}, {Gorjian}, {Bhattacharya}, {Carey}, {Nelson},
  {Glaccum}, {Lacy}, {Lowrance}, {Laine}, {Reach}, {Stauffer}, {Surace},
  {Wilson}, {Wright}, {Hoffman}, {Domingo}, \& {Cohen}}]{Fazio2004}
{Fazio}, G.~G., {Hora}, J.~L., {Allen}, L.~E., {et~al.} 2004, \apjs, 154, 10,
  \dodoi{10.1086/422843}

\bibitem[{{Feldman} {et~al.}(2004){Feldman}, {Cochran}, \&
  {Combi}}]{Feldman2004}
{Feldman}, P.~D., {Cochran}, A.~L., \& {Combi}, M.~R. 2004, {Spectroscopic
  investigations of fragment species in the coma}, ed. M.~C. {Festou}, H.~U.
  {Keller}, \& H.~A. {Weaver}, 425

\bibitem[{{Fern{\'a}ndez} {et~al.}(1999){Fern{\'a}ndez}, {Tancredi}, {Rickman},
  \& {Licandro}}]{Fernandez1999aap}
{Fern{\'a}ndez}, J.~A., {Tancredi}, G., {Rickman}, H., \& {Licandro}, J. 1999,
  \aap, 352, 327

\bibitem[{{Fern{\'a}ndez}(2009)}]{Fernandez2009}
{Fern{\'a}ndez}, Y.~R. 2009, \planss, 57, 1218,
  \dodoi{10.1016/j.pss.2009.01.003}

\bibitem[{{Fern{\'a}ndez} {et~al.}(2013){Fern{\'a}ndez}, {Kelley}, {Lamy},
  {Toth}, {Groussin}, {Lisse}, {A'Hearn}, {Bauer}, {Campins}, {Fitzsimmons},
  {Licandro}, {Lowry}, {Meech}, {Pittichov{\'a}}, {Reach}, {Snodgrass}, \&
  {Weaver}}]{Fernandez2013}
{Fern{\'a}ndez}, Y.~R., {Kelley}, M.~S., {Lamy}, P.~L., {et~al.} 2013, \icarus,
  226, 1138, \dodoi{10.1016/j.icarus.2013.07.021}

\bibitem[{{Fitzsimmons} {et~al.}(2020){Fitzsimmons}, {Young}, {Armstrong},
  {Moss}, \& {Sato}}]{Fitzsimmons2020MPEC}
{Fitzsimmons}, A., {Young}, D., {Armstrong}, J., {Moss}, S., \& {Sato}, H.
  2020, Minor Planet Electronic Circulars, 2020-K134, 1

\bibitem[{{Gladman} {et~al.}(2008){Gladman}, {Marsden}, \&
  {Vanlaerhoven}}]{Gladman2008}
{Gladman}, B., {Marsden}, B.~G., \& {Vanlaerhoven}, C. 2008, {Nomenclature in
  the Outer Solar System}, ed. M.~A. {Barucci}, H.~{Boehnhardt}, D.~P.
  {Cruikshank}, A.~{Morbidelli}, \& R.~{Dotson}, 43--57

\bibitem[{{Graham} {et~al.}(2019){Graham}, {Kulkarni}, {Bellm}, {Adams},
  {Barbarino}, {Blagorodnova}, {Bodewits}, {Bolin}, {Brady}, {Cenko}, {Chang},
  {Coughlin}, \& {De}}]{Graham2019}
{Graham}, M.~J., {Kulkarni}, S.~R., {Bellm}, E.~C., {et~al.} 2019, \pasp, 131,
  078001, \dodoi{10.1088/1538-3873/ab006c}

\bibitem[{{Gunnarsson} {et~al.}(2008){Gunnarsson}, {Bockel{\'e}e-Morvan},
  {Biver}, {Crovisier}, \& {Rickman}}]{Gunnarsson2008ada}
{Gunnarsson}, M., {Bockel{\'e}e-Morvan}, D., {Biver}, N., {Crovisier}, J., \&
  {Rickman}, H. 2008, \aap, 484, 537, \dodoi{10.1051/0004-6361:20078069}

\bibitem[{{Hanu{\v s}} {et~al.}(2018){Hanu{\v s}}, {Delbo}, {Al{\'{\i}}-Lagoa},
  {Bolin}, {Jedicke}, {{\v D}urech}, {Cibulkov{\'a}}, {Pravec}, {Ku{\v
  s}nir{\'a}k}, {Behrend}, {Marchis}, {Antonini}, {Arnold}, {Audejean},
  {Bachschmidt}, {Bernasconi}, {Brunetto}, {Casulli}, {Dymock}, {Esseiva},
  {Esteban}, {Gerteis}, {de Groot}, {Gully}, {Hamanowa}, {Hamanowa}, {Krafft},
  {Lehk{\'y}}, {Manzini}, {Michelet}, {Morelle}, {Oey}, {Pilcher}, {Reignier},
  {Roy}, {Salom}, \& {Warner}}]{Hanus2018}
{Hanu{\v s}}, J., {Delbo}, M., {Al{\'{\i}}-Lagoa}, V., {et~al.} 2018, \icarus,
  299, 84, \dodoi{10.1016/j.icarus.2017.07.007}

\bibitem[{{Haser}(1957)}]{Haser1957}
{Haser}, L. 1957, Bulletin de la Societe Royale des Sciences de Liege, 43, 740

\bibitem[{{Horner} \& {Evans}(2006)}]{Horner2006}
{Horner}, J., \& {Evans}, N.~W. 2006, \mnras, 367, L20,
  \dodoi{10.1111/j.1745-3933.2006.00131.x}

\bibitem[{Horner {et~al.}(2004)Horner, Evans, \& Bailey}]{Horner2004CentaursI}
Horner, J., Evans, N.~W., \& Bailey, M.~E. 2004, Mon. Not. R. Astron. Soc.,
  354, 798, \dodoi{10.1111/j.1365-2966.2004.08240.x}

\bibitem[{{Horner} {et~al.}(2004){Horner}, {Evans}, \& {Bailey}}]{Horner2004b}
{Horner}, J., {Evans}, N.~W., \& {Bailey}, M.~E. 2004, \mnras, 355, 321,
  \dodoi{10.1111/j.1365-2966.2004.08342.x}

\bibitem[{{Horner} \& {Lykawka}(2012)}]{Horner2012}
{Horner}, J., \& {Lykawka}, P.~S. 2012, \mnras, 426, 159,
  \dodoi{10.1111/j.1365-2966.2012.21717.x}

\bibitem[{Hsieh {et~al.}(2021)Hsieh, Fitzsimmons, Novakovi{\'{c}}, Denneau, \&
  Heinze}]{Hsieh20202019ld2}
Hsieh, H.~H., Fitzsimmons, A., Novakovi{\'{c}}, B., Denneau, L., \& Heinze,
  A.~N. 2021, Icarus, 354, 114019, \dodoi{10.1016/j.icarus.2020.114019}

\bibitem[{{Hsieh} {et~al.}(2015{\natexlab{a}}){Hsieh}, {Denneau}, {Wainscoat},
  {Sch{\"o}rghofer}, {Bolin}, {Fitzsimmons}, {Jedicke}, {Kleyna}, {Micheli},
  {Vere{\v s}}, {Kaiser}, {Chambers}, {Burgett}, {Flewelling}, {Hodapp},
  {Magnier}, {Morgan}, {Price}, {Tonry}, \& {Waters}}]{Hsieh2015}
{Hsieh}, H.~H., {Denneau}, L., {Wainscoat}, R.~J., {et~al.} 2015{\natexlab{a}},
  \icarus, 248, 289, \dodoi{10.1016/j.icarus.2014.10.031}

\bibitem[{{Hsieh} {et~al.}(2015{\natexlab{b}}){Hsieh}, {Hainaut},
  {Novakovi{\'c}}, {Bolin}, {Denneau}, {Fitzsimmons}, {Haghighipour}, {Kleyna},
  {Kokotanekova}, {Lacerda}, {Meech}, {Micheli}, {Moskovitz}, {Schunova},
  {Snodgrass}, {Wainscoat}, {Wasserman}, \& {Waszczak}}]{Hsieh2015a}
{Hsieh}, H.~H., {Hainaut}, O., {Novakovi{\'c}}, B., {et~al.}
  2015{\natexlab{b}}, \apjl, 800, L16, \dodoi{10.1088/2041-8205/800/1/L16}

\bibitem[{{Huehnerhoff} {et~al.}(2016){Huehnerhoff}, {Ketzeback}, {Bradley},
  {Dembicky}, {Doughty}, {Hawley}, {Johnson}, {Klaene}, {Leon}, {McMillan},
  {Owen}, {Sayres}, {Sheen}, \& {Shugart}}]{Huehnerhoff2016}
{Huehnerhoff}, J., {Ketzeback}, W., {Bradley}, A., {et~al.} 2016, in \procspie,
  Vol. 9908, Ground-based and Airborne Instrumentation for Astronomy VI, 99085H

\bibitem[{{Ivanova} {et~al.}(2014){Ivanova}, {Borysenko}, \&
  {Golovin}}]{Ivanova2014}
{Ivanova}, O., {Borysenko}, S., \& {Golovin}, A. 2014, \icarus, 227, 202,
  \dodoi{10.1016/j.icarus.2013.08.026}

\bibitem[{{Ivanova} {et~al.}(2011){Ivanova}, {Skorov}, {Korsun}, {Afanasiev},
  \& {Blum}}]{Ivanova2011}
{Ivanova}, O.~V., {Skorov}, Y.~V., {Korsun}, P.~P., {Afanasiev}, V.~L., \&
  {Blum}, J. 2011, \icarus, 211, 559, \dodoi{10.1016/j.icarus.2010.10.026}

\bibitem[{{Ivezi{\'c}} {et~al.}(2001){Ivezi{\'c}}, {Tabachnik}, {Rafikov},
  {Lupton}, {Quinn}, {Hammergren}, {Eyer}, {Chu}, {Armstrong}, {Fan},
  {Finlator}, {Geballe}, {Gunn}, {Hennessy}, {Knapp}, {Leggett}, {Munn},
  {Pier}, {Rockosi}, {Schneider}, {Strauss}, {Yanny}, {Brinkmann}, {Csabai},
  {Hindsley}, {Kent}, {Lamb}, {Margon}, {McKay}, {Smith}, {Waddel}, {York}, \&
  {SDSS Collaboration}}]{Ivezic2001}
{Ivezi{\'c}}, {\v Z}., {Tabachnik}, S., {Rafikov}, R., {et~al.} 2001, \aj, 122,
  2749, \dodoi{10.1086/323452}

\bibitem[{{Ivezi{\'c}} {et~al.}(2002){Ivezi{\'c}}, {Lupton}, {Juri{\'c}},
  {Tabachnik}, {Quinn}, {Gunn}, {Knapp}, {Rockosi}, \&
  {Brinkmann}}]{Ivezic2002}
{Ivezi{\'c}}, {\v Z}., {Lupton}, R.~H., {Juri{\'c}}, M., {et~al.} 2002, \aj,
  124, 2943, \dodoi{10.1086/344077}

\bibitem[{{Jedicke} {et~al.}(2016){Jedicke}, {Bolin}, {Granvik}, \&
  {Beshore}}]{Jedicke2016}
{Jedicke}, R., {Bolin}, B., {Granvik}, M., \& {Beshore}, E. 2016, \icarus, 266,
  173, \dodoi{10.1016/j.icarus.2015.10.021}

\bibitem[{{Jewitt}(2015)}]{Jewitt2015bbb}
{Jewitt}, D. 2015, \aj, 150, 201, \dodoi{10.1088/0004-6256/150/6/201}

\bibitem[{{Jewitt} {et~al.}(2015){Jewitt}, {Hsieh}, \& {Agarwal}}]{Jewitt2015a}
{Jewitt}, D., {Hsieh}, H., \& {Agarwal}, J. 2015, {The Active Asteroids}, ed.
  P.~{Michel}, F.~E. {DeMeo}, \& W.~F. {Bottke}, 221--241

\bibitem[{{Jewitt} {et~al.}(2014){Jewitt}, {Ishiguro}, {Weaver}, {Agarwal},
  {Mutchler}, \& {Larson}}]{Jewitt2014}
{Jewitt}, D., {Ishiguro}, M., {Weaver}, H., {et~al.} 2014, \aj, 147, 117,
  \dodoi{10.1088/0004-6256/147/5/117}

\bibitem[{{Jewitt}(2009)}]{Jewitt2009}
{Jewitt}, D.~C. 2009, in European Planetary Science Congress 2009, 13

\bibitem[{{Jewitt} \& {Meech}(1987)}]{Jewitt1987ab}
{Jewitt}, D.~C., \& {Meech}, K.~J. 1987, \apj, 317, 992, \dodoi{10.1086/165347}

\bibitem[{{Jordi} {et~al.}(2006){Jordi}, {Grebel}, \& {Ammon}}]{Jordi2006}
{Jordi}, K., {Grebel}, E.~K., \& {Ammon}, K. 2006, \aap, 460, 339,
  \dodoi{10.1051/0004-6361:20066082}

\bibitem[{{Juri{\'c}} {et~al.}(2002){Juri{\'c}}, {Ivezi{\'c}}, {Lupton},
  {Quinn}, {Tabachnik}, {Fan}, {Gunn}, {Hennessy}, {Knapp}, {Munn}, {Pier},
  {Rockosi}, {Schneider}, {Brinkmann}, {Csabai}, \& {Fukugita}}]{Juric2002}
{Juri{\'c}}, M., {Ivezi{\'c}}, {\v Z}., {Lupton}, R.~H., {et~al.} 2002, \aj,
  124, 1776, \dodoi{10.1086/341950}

\bibitem[{{Kareta} {et~al.}(2020{\natexlab{a}}){Kareta}, {Volk}, {Noonan},
  {Sharkey}, {Harris}, \& {Reddy}}]{Kareta2020RNAAS}
{Kareta}, T., {Volk}, K., {Noonan}, J.~W., {et~al.} 2020{\natexlab{a}},
  Research Notes of the American Astronomical Society, 4, 74,
  \dodoi{10.3847/2515-5172/ab963c}

\bibitem[{{Kareta} {et~al.}(2020{\natexlab{b}}){Kareta}, {Woodney},
  {Schambeau}, {Fernandez}, {Harrington Pinto}, {Wierzchos}, {Womack}, {Bus},
  {Steckloff}, {Sarid}, {Volk}, {Harris}, \& {Reddy}}]{Kareta2020ld2}
{Kareta}, T., {Woodney}, L.~M., {Schambeau}, C., {et~al.} 2020{\natexlab{b}},
  arXiv e-prints, arXiv:2011.09993.
\newblock \doarXiv{2011.09993}

\bibitem[{{Kasliwal} {et~al.}(2019){Kasliwal}, {Cannella}, {Bagdasaryan},
  {Hung}, {Feindt}, {Singer}, {Coughlin}, {Fremling}, {Walters}, {Duev},
  {Itoh}, \& {Quimby}}]{Kasliwal2019}
{Kasliwal}, M.~M., {Cannella}, C., {Bagdasaryan}, A., {et~al.} 2019, \pasp,
  131, 038003, \dodoi{10.1088/1538-3873/aafbc2}

\bibitem[{{Kelley} {et~al.}(2013){Kelley}, {Fern{\'a}ndez}, {Licandro},
  {Lisse}, {Reach}, {A'Hearn}, {Bauer}, {Campins}, {Fitzsimmons}, {Groussin},
  {Lamy}, {Lowry}, {Meech}, {Pittichov{\'a}}, {Snodgrass}, {Toth}, \&
  {Weaver}}]{Kelley2013}
{Kelley}, M.~S., {Fern{\'a}ndez}, Y.~R., {Licandro}, J., {et~al.} 2013,
  \icarus, 225, 475, \dodoi{10.1016/j.icarus.2013.04.012}

\bibitem[{{Kelley} {et~al.}(2019){Kelley}, {Bodewits}, {Ye}, {Laher}, {Masci},
  {Monkewitz}, {Riddle}, {Rusholme}, {Shupe}, \& {Soumagnac}}]{Kelley2019}
{Kelley}, M. S.~P., {Bodewits}, D., {Ye}, Q., {et~al.} 2019, in Astronomical
  Society of the Pacific Conference Series, Vol. 523, Astronomical Data
  Analysis Software and Systems XXVII, ed. P.~J. {Teuben}, M.~W. {Pound}, B.~A.
  {Thomas}, \& E.~M. {Warner}, 471

\bibitem[{{Kinoshita} {et~al.}(2005){Kinoshita}, {Chen}, {Lin}, {Lin}, {Huang},
  {Chang}, \& {Chen}}]{Kinoshita2005}
{Kinoshita}, D., {Chen}, C.-W., {Lin}, H.-C., {et~al.} 2005, \cjaa, 5, 315,
  \dodoi{10.1088/1009-9271/5/3/011}

\bibitem[{{Krist} {et~al.}(2011){Krist}, {Hook}, \& {Stoehr}}]{Krist2011}
{Krist}, J.~E., {Hook}, R.~N., \& {Stoehr}, F. 2011, in Society of
  Photo-Optical Instrumentation Engineers (SPIE) Conference Series, Vol. 8127,
  Optical Modeling and Performance Predictions V, 81270J

\bibitem[{{L{\"a}uter} {et~al.}(2019){L{\"a}uter}, {Kramer}, {Rubin}, \&
  {Altwegg}}]{Lauter2019}
{L{\"a}uter}, M., {Kramer}, T., {Rubin}, M., \& {Altwegg}, K. 2019, \mnras,
  483, 852, \dodoi{10.1093/mnras/sty3103}

\bibitem[{{Levison} \& {Duncan}(1994)}]{Levison1994}
{Levison}, H.~F., \& {Duncan}, M.~J. 1994, \icarus, 108, 18,
  \dodoi{10.1006/icar.1994.1039}

\bibitem[{{Licandro} {et~al.}(2020){Licandro}, {Pinilla-Alonso}, {de Leon},
  {Moreno}, {de la Fuente Marcos}, {de la Fuente Marcos}, \& {Cabrera
  Lavers}}]{Licandro2020DPS}
{Licandro}, J., {Pinilla-Alonso}, N., {de Leon}, J., {et~al.} 2020, in
  AAS/Division for Planetary Sciences Meeting Abstracts, Vol.~52, AAS/Division
  for Planetary Sciences Meeting Abstracts, 404.06

\bibitem[{{Lin} {et~al.}(2020){Lin}, {Cheng}, {Vincent}, {Zhang}, {Ip}, \&
  {Chi}}]{Lin2020gault}
{Lin}, Z.-Y., {Cheng}, Y.-L., {Vincent}, J.-B., {et~al.} 2020, \pasj,
  \dodoi{10.1093/pasj/psaa069}

\bibitem[{Lisse {et~al.}(2020)Lisse, Bauer, Cruikshank, Emery, Fern{\'a}ndez,
  Fern{\'a}ndez-Valenzuela, Kelley, McKay, Reach, Pendleton, Pinilla-Alonso,
  Stansberry, Sykes, Trilling, Wooden, Harker, Gehrz, \&
  Woodward}]{Lisse2020Spitzer}
Lisse, C., Bauer, J., Cruikshank, D., {et~al.} 2020, Nature Astronomy, 4, 930,
  \dodoi{10.1038/s41550-020-01219-6}

\bibitem[{{Lisse} {et~al.}(2019){Lisse}, {Young}, {Cruikshank}, {Sandford}, \&
  {Schmitt}}]{Lisse2019}
{Lisse}, C.~M., {Young}, L.~A., {Cruikshank}, D., {Sandford}, S., \& {Schmitt},
  B. 2019, Submitted to Icarus

\bibitem[{{Lisse} {et~al.}(2020){Lisse}, {Young}, {Cruikshank}, {Sand ford},
  {Schmitt}, {Stern}, {Weaver}, {Umurhan}, {Pendleton}, {Keane}, {Gladstone},
  {Parker}, {Binzel}, {Earle}, {Horanyi}, {El-Maarry}, {Cheng}, {Moore},
  {McKinnon}, {Grundy}, {Kavelaars}, {Linscott}, {Lyra}, {Lewis}, {Britt},
  {Spencer}, {Olkin}, {McNutt}, {Elliott}, {Dello-Russo}, {Steckloff}, {Neveu},
  \& {Mousis}}]{Lisse2020}
{Lisse}, C.~M., {Young}, L.~A., {Cruikshank}, D.~P., {et~al.} 2020, arXiv
  e-prints, arXiv:2009.02277.
\newblock \doarXiv{2009.02277}

\bibitem[{{Marshall} {et~al.}(2019){Marshall}, {Rezac}, {Hartogh}, {Zhao}, \&
  {Attree}}]{Marshall2019}
{Marshall}, D., {Rezac}, L., {Hartogh}, P., {Zhao}, Y., \& {Attree}, N. 2019,
  \aap, 623, A120, \dodoi{10.1051/0004-6361/201833959}

\bibitem[{{Marzari} {et~al.}(2002){Marzari}, {Scholl}, {Murray}, \&
  {Lagerkvist}}]{Marzari2002}
{Marzari}, F., {Scholl}, H., {Murray}, C., \& {Lagerkvist}, C. 2002, {Origin
  and Evolution of Trojan Asteroids}, 725--738

\bibitem[{{Masci} {et~al.}(2019){Masci}, {Laher}, {Rusholme}, {Shupe}, {Groom},
  {Surace}, {Jackson}, {Monkewitz}, {Beck}, {Flynn}, {Terek}, {Landry},
  {Hacopians}, {Desai}, {Howell}, {Brooke}, {Imel}, {Wachter}, {Ye}, {Lin},
  {Cenko}, {Cunningham}, {Rebbapragada}, {Bue}, {Miller}, {Mahabal}, {Bellm},
  {Patterson}, {Juri{\'c}}, {Golkhou}, {Ofek}, {Walters}, {Graham}, {Kasliwal},
  {Dekany}, {Kupfer}, {Burdge}, {Cannella}, {Barlow}, {Van Sistine}, {Giomi},
  {Fremling}, {Blagorodnova}, {Levitan}, {Riddle}, {Smith}, {Helou}, {Prince},
  \& {Kulkarni}}]{Masci2019}
{Masci}, F.~J., {Laher}, R.~R., {Rusholme}, B., {et~al.} 2019, \pasp, 131,
  018003, \dodoi{10.1088/1538-3873/aae8ac}

\bibitem[{{McCarthy} {et~al.}(1998){McCarthy}, {Cohen}, {Butcher}, {Cromer},
  {Croner}, {Douglas}, {Goeden}, {Grewal}, {Lu}, {Petrie}, {Weng}, {Weber},
  {Koch}, \& {Rodgers}}]{McCarthy1998}
{McCarthy}, J.~K., {Cohen}, J.~G., {Butcher}, B., {et~al.} 1998, Society of
  Photo-Optical Instrumentation Engineers (SPIE) Conference Series, Vol. 3355,
  {Blue channel of the Keck low-resolution imaging spectrometer}, ed.
  S.~{D'Odorico}, 81--92

\bibitem[{{McDonnell} {et~al.}(1986){McDonnell}, {Alexander}, {Burton},
  {Bussoletti}, {Clark}, {Grard}, {Gr{\"u}n}, {Hanner}, {Hughes}, {Igenbergs},
  {Kuczera}, {Lindblad}, {Mand eville}, {Minafra}, {Schwehm}, {Sekanina},
  {Wallis}, {Zarnecki}, {Chakaveh}, {Evans}, {Evans}, {Firth}, {Littler},
  {Massonne}, {Olearczyk}, {Pankiewicz}, {Stevenson}, \&
  {Turner}}]{McDonnell1986}
{McDonnell}, J.~A.~M., {Alexander}, W.~M., {Burton}, W.~M., {et~al.} 1986,
  \nat, 321, 338, \dodoi{10.1038/321338a0}

\bibitem[{{Meech} \& {Svoren}(2004)}]{Meech2004}
{Meech}, K.~J., \& {Svoren}, J. 2004, {Using cometary activity to trace the
  physical and chemical evolution of cometary nuclei}, ed. M.~C. {Festou},
  H.~U. {Keller}, \& H.~A. {Weaver}, 317

\bibitem[{{Mommert} {et~al.}(2019){Mommert}, {Kelley}, {de Val-Borro}, {Li},
  {Guzman}, {Sip{\H{o}}cz}, {{\v{D}}urech}, {Granvik}, {Grundy}, {Moskovitz},
  {Penttil{\"a}}, \& {Samarasinha}}]{Mommert2019ab}
{Mommert}, M., {Kelley}, M., {de Val-Borro}, M., {et~al.} 2019, The Journal of
  Open Source Software, 4, 1426, \dodoi{10.21105/joss.01426}

\bibitem[{{Morbidelli} {et~al.}(2005){Morbidelli}, {Levison}, {Tsiganis}, \&
  {Gomes}}]{Morbidelli2005}
{Morbidelli}, A., {Levison}, H.~F., {Tsiganis}, K., \& {Gomes}, R. 2005, \nat,
  435, 462, \dodoi{10.1038/nature03540}

\bibitem[{{Moreno} {et~al.}(2017){Moreno}, {Pozuelos}, {Novakovi{\'c}},
  {Licandro}, {Cabrera-Lavers}, {Bolin}, {Jedicke}, {Gladman}, {Bannister},
  {Gwyn}, {Vere{\v s}}, {Chambers}, {Chastel}, {Denneau}, {Flewelling},
  {Huber}, {Schunov{\'a}-Lilly}, {Magnier}, {Wainscoat}, {Waters}, {Weryk},
  {Farnocchia}, \& {Micheli}}]{Moreno2017}
{Moreno}, F., {Pozuelos}, F.~J., {Novakovi{\'c}}, B., {et~al.} 2017, \apjl,
  837, L3, \dodoi{10.3847/2041-8213/aa6036}

\bibitem[{{Murray} \& {Dermott}(1999)}]{Murray1999}
{Murray}, C.~D., \& {Dermott}, S.~F. 1999, {Solar system dynamics}

\bibitem[{{Nesvorn{\'y}} {et~al.}(2018){Nesvorn{\'y}}, {Parker}, \&
  {Vokrouhlick{\'y}}}]{Nesvorny2018}
{Nesvorn{\'y}}, D., {Parker}, J., \& {Vokrouhlick{\'y}}, D. 2018, \aj, 155,
  246, \dodoi{10.3847/1538-3881/aac01f}

\bibitem[{{Nesvorn{\'y}} {et~al.}(2017){Nesvorn{\'y}}, {Vokrouhlick{\'y}},
  {Dones}, {Levison}, {Kaib}, \& {Morbidelli}}]{Nesvorny2017}
{Nesvorn{\'y}}, D., {Vokrouhlick{\'y}}, D., {Dones}, L., {et~al.} 2017, \apj,
  845, 27, \dodoi{10.3847/1538-4357/aa7cf6}

\bibitem[{Ofek(2012)}]{Ofek2012}
Ofek, E.~O. 2012, The Astrophysical Journal, 749, 10

\bibitem[{{Oke} {et~al.}(1995){Oke}, {Cohen}, {Carr}, {Cromer}, {Dingizian},
  {Harris}, {Labrecque}, {Lucinio}, {Schaal}, {Epps}, \& {Miller}}]{Oke1995}
{Oke}, J.~B., {Cohen}, J.~G., {Carr}, M., {et~al.} 1995, \pasp, 107, 375,
  \dodoi{10.1086/133562}

\bibitem[{O'Rourke {et~al.}(2020)O'Rourke, Heinisch, Blum, Fornasier,
  Filacchione, Van~Hoang, Ciarniello, Raponi, Gundlach, Blasco, Grieger,
  Glassmeier, K{\"u}ppers, Rotundi, Groussin, Bockel{\'e}e-Morvan, Auster,
  Oklay, Paar, Perucha, Kovacs, Jorda, Vincent, Capaccioni, Biver, Parker,
  Tubiana, \& Sierks}]{ORourke2020}
O'Rourke, L., Heinisch, P., Blum, J., {et~al.} 2020, Nature, 586, 697,
  \dodoi{10.1038/s41586-020-2834-3}

\bibitem[{{Perley}(2019)}]{Perley2019}
{Perley}, D.~A. 2019, \pasp, 131, 084503, \dodoi{10.1088/1538-3873/ab215d}

\bibitem[{{Reach} {et~al.}(2013){Reach}, {Kelley}, \& {Vaubaillon}}]{Reach2013}
{Reach}, W.~T., {Kelley}, M.~S., \& {Vaubaillon}, J. 2013, \icarus, 226, 777,
  \dodoi{10.1016/j.icarus.2013.06.011}

\bibitem[{Rein \& Liu(2012)}]{Rein2012REBOUND}
Rein, H., \& Liu, S.-F. 2012, Astron. Astrophys., 537, A128,
  \dodoi{10.1051/0004-6361/201118085}

\bibitem[{Rein {et~al.}(2019)Rein, Hernandez, Tamayo, Brown, Eckels, Holmes,
  Lau, Leblanc, \& Silburt}]{Rein2019MERCURIUS}
Rein, H., Hernandez, D.~M., Tamayo, D., {et~al.} 2019, Mon. Not. R. Astron.
  Soc., 485, 5490, \dodoi{10.1093/mnras/stz769}

\bibitem[{{Roig} \& {Nesvorn{\'y}}(2015)}]{Roig2015}
{Roig}, F., \& {Nesvorn{\'y}}, D. 2015, \aj, 150, 186,
  \dodoi{10.1088/0004-6256/150/6/186}

\bibitem[{{Sarid} \& {Prialnik}(2009)}]{Sarid2009}
{Sarid}, G., \& {Prialnik}, D. 2009, Meteoritics and Planetary Science, 44,
  1905, \dodoi{10.1111/j.1945-5100.2009.tb02000.x}

\bibitem[{{Sarid} {et~al.}(2019){Sarid}, {Volk}, {Steckloff}, {Harris},
  {Womack}, \& {Woodney}}]{Sarid2019}
{Sarid}, G., {Volk}, K., {Steckloff}, J.~K., {et~al.} 2019, \apjl, 883, L25,
  \dodoi{10.3847/2041-8213/ab3fb3}

\bibitem[{{Sato} {et~al.}(2020){Sato}, {Heinze}, \& {Nakano}}]{Sato2020CBET}
{Sato}, H., {Heinze}, A., \& {Nakano}, S. 2020, Central Bureau Electronic
  Telegrams, 4780, 1

\bibitem[{{Schambeau} {et~al.}(2020){Schambeau}, {Fernandez}, {Belton},
  {Womack}, {Micheli}, {Woodney}, {Steckloff}, {Sarid}, {Kareta}, {Harris}, \&
  {Volk}}]{Schambeau2020cbet}
{Schambeau}, C., {Fernandez}, Y., {Belton}, R., {et~al.} 2020, Central Bureau
  Electronic Telegrams, 4821, 1

\bibitem[{{Schambeau} {et~al.}(2019){Schambeau}, {Fern{\'a}ndez},
  {Samarasinha}, {Woodney}, \& {Kundu}}]{Schambeau2019}
{Schambeau}, C.~A., {Fern{\'a}ndez}, Y.~R., {Samarasinha}, N.~H., {Woodney},
  L.~M., \& {Kundu}, A. 2019, \aj, 158, 259, \dodoi{10.3847/1538-3881/ab53e2}

\bibitem[{{Schleicher} \& {Bair}(2011)}]{Schleicher2011}
{Schleicher}, D.~G., \& {Bair}, A.~N. 2011, \aj, 141, 177,
  \dodoi{10.1088/0004-6256/141/6/177}

\bibitem[{{Smith} \& {Nelson}(1969)}]{Smith_Nelson1969}
{Smith}, C.~E., \& {Nelson}, B. 1969, \pasp, 81, 74, \dodoi{10.1086/128742}

\bibitem[{{Solontoi} {et~al.}(2012){Solontoi}, {Ivezi{\'c}}, {Juri{\'c}},
  {Becker}, {Jones}, {West}, {Kent}, {Lupton}, {Claire}, {Knapp}, {Quinn},
  {Gunn}, \& {Schneider}}]{Solontoi2012}
{Solontoi}, M., {Ivezi{\'c}}, {\v Z}., {Juri{\'c}}, M., {et~al.} 2012, \icarus,
  218, 571, \dodoi{10.1016/j.icarus.2011.10.008}

\bibitem[{{Steckloff} {et~al.}(2020){Steckloff}, {Sarid}, {Volk}, {Kareta},
  {Womack}, {Harris}, {Woodney}, \& {Schambeau}}]{Steckloff2020}
{Steckloff}, J.~K., {Sarid}, G., {Volk}, K., {et~al.} 2020, \apjl, 904, L20,
  \dodoi{10.3847/2041-8213/abc888}

\bibitem[{Steele {et~al.}(2004)Steele, Smith, Rees, Baker, Bates, Bode, Bowman,
  Carter, Etherton, Ford, Fraser, Gomboc, Lett, Mansfield, Marchant,
  Medrano-Cerda, Mottram, Raback, Scott, Tomlinson, \& Zamanov}]{Steele2004}
Steele, I.~A., Smith, R.~J., Rees, P. C.~T., {et~al.} 2004, in SPIE
  Astronomical Telescopes + Instrumentation

\bibitem[{{Tonry}(2011)}]{Tonry2011}
{Tonry}, J.~L. 2011, \pasp, 123, 58, \dodoi{10.1086/657997}

\bibitem[{{Vokrouhlick{\'y}} {et~al.}(2019){Vokrouhlick{\'y}}, {Nesvorn{\'y}},
  \& {Dones}}]{Vokrouhlicky2019}
{Vokrouhlick{\'y}}, D., {Nesvorn{\'y}}, D., \& {Dones}, L. 2019, \aj, 157, 181,
  \dodoi{10.3847/1538-3881/ab13aa}

\bibitem[{{Vokrouhlick{\'y}} {et~al.}(2017){Vokrouhlick{\'y}}, {Pravec},
  {Durech}, {Bolin}, {Jedicke}, {Ku{\v s}nir{\'a}k}, {Gal{\'a}d}, {Hornoch},
  {Kryszczy{\'n}ska}, {Colas}, {Moskovitz}, {Thirouin}, \&
  {Nesvorn{\'y}}}]{Vokrouhlicky2017a}
{Vokrouhlick{\'y}}, D., {Pravec}, P., {Durech}, J., {et~al.} 2017, \aap, 598,
  A91, \dodoi{10.1051/0004-6361/201629670}

\bibitem[{{Werner} {et~al.}(2004){Werner}, {Roellig}, {Low}, {Rieke}, {Rieke},
  {Hoffmann}, {Young}, {Houck}, {Brandl}, {Fazio}, {Hora}, {Gehrz}, {Helou},
  {Soifer}, {Stauffer}, {Keene}, {Eisenhardt}, {Gallagher}, {Gautier}, {Irace},
  {Lawrence}, {Simmons}, {Van Cleve}, {Jura}, {Wright}, \&
  {Cruikshank}}]{Werner2004}
{Werner}, M.~W., {Roellig}, T.~L., {Low}, F.~J., {et~al.} 2004, \apjs, 154, 1,
  \dodoi{10.1086/422992}

\bibitem[{{Ye} {et~al.}(2019){Ye}, {Kelley}, {Bodewits}, {Bolin}, {Jones},
  {Lin}, {Bellm}, {Dekany}, {Duev}, {Groom}, {Helou}, {Kulkarni}, {Kupfer},
  {Masci}, {Prince}, \& {Soumagnac}}]{Ye2019adfs}
{Ye}, Q., {Kelley}, M. S.~P., {Bodewits}, D., {et~al.} 2019, \apjl, 874, L16,
  \dodoi{10.3847/2041-8213/ab0f3c}

\end{thebibliography}

\newpage
\setcounter{footnote}{0}
\renewcommand{\thefootnote}{\Roman{footnote}}

\begin{longtable}{|l|c|c|c|c|c|c|c|c|c|}
\caption{Summary of \ld target observations viewing geometry.\label{tab:obs}}\\
\hline
Date$^1$ & Facility$^2$ & Filter$^3$& $\theta_s^4$&$\chi_{am}^{5}$&$r_H^6$&$\Delta^7$&$\alpha^8$ & $\delta_{E}^{9}$ & $T \, - \, T_p^{10}$\\
UTC&&&($\arcsec$)&&(au)&(au)&($^{\circ}$)&($^{\circ}$)&(days)\\
\hline
\endfirsthead
\multicolumn{4}{c}%
{\tablename\ \thetable\ -- \textit{Continued from previous page}} \\
\hline
Date$^1$ & Facility$^2$ & Filter$^3$& $\theta_s^4$&$\chi_{am}^{5}$&$r_H^6$&$\Delta^7$&$\alpha^8$ & $\delta_{E}^{9}$ & $T \, - \, T_p^{10}$\\
UTC&&&($\arcsec$)&&(au)&(au)&($^{\circ}$)&($^{\circ}$)&(days)\\
\hline
\endhead
\hline \multicolumn{4}{r}{\textit{Continued on next page}} \\
\endfoot
\hline
\endlastfoot
2019 Apr 26 & ZTF\footnote[1]{Serendipitous observation of \ld with ZTF, but included in this table because of its inclusion in the top-left panel of Fig.~\ref{fFiig:mosaic}.} & $r$ & 2.17 &1.76 & 4.693 & 4.147 & 10.99 & -1.56 & -343.97 \\
2019 Sep 07 & ARC & $g$,$r$ & 1.4 &1.81  & 4.622 & 4.279 & 12.23 & -0.74 & -216.58 \\
2020 Jan 25-26 & \textit{Spitzer} & 4.5 $\mu$m & $-$ &$-$ & 4.584 & 4.256\footnote[2]{\label{note2}\textit{Spitzer}-centric} & 12.61$\mathrm{^{\RomanNumeralCaps{2}}}$ & 0.23$\mathrm{^{\RomanNumeralCaps{2}}}$ & -76.58 \\
2020 Apr 01 & \textit{HST} & F350LP & $-$ &$-$ & 4.578 & 5.023 & 10.71 & -0.44 & -9.58 \\
2020 May 27 & MLO 1.0-m & $B$,$V$,$R$ & 1.92 &1.49 & 4.580 & 4.221 & 12.38 & -2.50 & 46.42 \\
2020 May 29 & LT & $g$,$r$,$i$,$z$ & 1.21 &1.75 & 4.580 & 4.193 & 12.27  & -2.56 & 48.42 \\
2020 June 23-27 & LOT & $B$,$V$,$R$ & 1.47&1.14  & 4.583 & 3.848 & 9.59  & -3.00 & 75.42 \\
2020 July 10 & LOT & $B$,$V$,$R$ & 1.45 &1.15 & 4.586 & 3.707 & 7.11  & -2.97 & 90.42\\
2020 Aug 19 & Keck I & Sp.\footnote[3]{\label{note1}\texttt{https://www2.keck.hawaii.edu/inst/lris/filters.html}} & 0.85 &1.80 & 4.594 & 3.608 & 3.20 & -1.76 & 130.42\\
\hline
\caption{Columns: (1) observation date; (2) observational facility; (3) filter (for the Keck I observations, the B600/4000 grism and R600/7500 grating are used with the LRIS instrument\footnote[2]{\ref{note1}}
; (4) in-image seeing of observations; (5) airmass of observations; (6) heliocentric distance; (7) topo-centric distance; (8) phase angle; (9) topo-centric and target orbital plane angle; (10) difference between time of observation $T$ and time of perihelion $T_p$}
\end{longtable}

\begin{longtable}{|l|c|c|c|c|c|c|c|c|c|c|c|}
\caption{Summary of ZTF \ld photometry.\label{tab:phot}}\\
\hline
Date$^1$ & $r_H^2$ & $\Delta^3$&$\alpha^4$&$T \, - \, T_p^5$&filter$^6$&mag$^7$& $\sigma_\mathrm{mag}^8$ & $\theta_s^9$ & $\chi_{am}^{10}$ & \afrozns$^{11}$  & C$^{12}$\\
UTC&(au)&(au)&($^{\circ}$)&(days)&&&&($\arcsec$)&&(cm)&(km$^2$)\\
\hline
\endfirsthead
\multicolumn{4}{c}%
{\tablename\ \thetable\ -- \textit{Continued from previous page}} \\
\hline
Date$^1$ & $r_H^2$ & $\Delta^3$&$\alpha^4$&$T \, - \, T_p^5$&filter$^6$&mag$^7$& $\sigma_\mathrm{mag}^8$ & $\theta_s^9$ & $\chi_{am}^{10}$ & \afrozns$^{11}$  & C$^{12}$\\
UTC&(au)&(au)&($^{\circ}$)&(days)&&&&($\arcsec$)&&(cm)&(km$^2$)\\
\hline
\endhead
\hline \multicolumn{4}{r}{\textit{Continued on next page}} \\
\endfoot
\hline
\endlastfoot
2019 Apr 09 - 10:32 & 4.704 & 4.391 & 12.0 & -360.20 & $r$ & 19.33 & 0.15 & 2.36& 1.72 & 85.28 & 127.69 \\
2019 Apr 12 - 10:32 & 4.702 & 4.345 & 11.9 & -357.32 & $r$ & 19.27 & 0.16 & 2.83& 1.66 & 87.90 & 131.53 \\
2019 Apr 15 - 09:59 & 4.700 & 4.300 & 11.7 & -354.47 & $r$ & 19.13 & 0.15 & 1.82 &2.02 & 97.23 & 145.35 \\
2019 Apr 20 - 09:33 & 4.697 & 4.227 & 11.4 & -349.70 & $r$ & 18.97 & 0.23 & 1.87 &1.83 & 107.69 & 160.77 \\
2019 Apr 20 - 10:04 & 4.697 & 4.226 & 11.4 & -349.68 & $r$ & 18.83 & 0.23 & 2.44 &1.51 & 122.45 & 182.81 \\
2019 Apr 26 - 09:00 & 4.693 & 4.142 & 11.0 & -343.97 & $g$ & 19.69 & 0.14 & 2.14 &1.91 & 83.20 & 124.00 \\
2019 Apr 26 - 09:07 & 4.693 & 4.142 & 11.0 & -343.96 & $g$ & 19.38 & 0.13 & 2.18 &2.12 & 110.69 & 164.97 \\
2019 Apr 26 - 11:26 & 4.693 & 4.141 & 11.0 & -343.87 & $r$ & 18.89 & 0.07 & 1.90 &1.41 & 109.63 & 163.38 \\
2019 Apr 26 - 11:36 & 4.693 & 4.141 & 11.0 & -343.86 & $r$ & 19.18 & 0.07 & 1.68 &1.37 & 83.93 & 125.08 \\
2019 May 02 - 10:03 & 4.689 & 4.061 & 10.4 & -338.17 & $r$ & 18.95 & 0.11 & 2.47 &1.56 & 97.64 & 145.18 \\
2019 May 02 - 10:32 & 4.689 & 4.061 & 10.4 & -338.15 & $r$ & 18.99 & 0.09 & 1.82 &1.41 & 94.11 & 139.93 \\
2019 May 31 - 08:22 & 4.672 & 3.763 & 6.1 & -310.33 & $r$ & 18.76 & 0.07 & 1.60 &1.44 & 85.34 & 125.82 \\
2019 May 31 - 10:34 & 4.672 & 3.762 & 6.1 & -310.24 & $g$ & 18.96 & 0.07 & 1.79& 1.44 & 112.44 & 165.78 \\
2019 Jun 01 - 09:04 & 4.671 & 3.755 & 6.0 & -309.33 & $r$ & 18.57 & 0.05 & 1.78 &1.38 & 100.81 & 148.64 \\
2019 Jun 02 - 09:34 & 4.671 & 3.748 & 5.8 & -308.35 & $r$ & 18.49 & 0.07 & 2.53 &1.37 & 107.32 & 158.24 \\
2019 Jun 02 - 10:57 & 4.671 & 3.748 & 5.8 & -308.29 & $g$ & 18.95 & 0.08 & 1.93& 1.52 & 111.35 & 164.18 \\
2019 Jun 03 - 11:33 & 4.670 & 3.741 & 5.6 & -307.30 & $g$ & 18.52 & 0.18 & 1.99 &1.71 & 163.56 & 241.16 \\
2019 Jun 04 - 07:01 & 4.670 & 3.736 & 5.5 & -306.52 & $r$ & 18.06 & 0.06 & 2.07 &1.67 & 156.64 & 230.96 \\
2019 Jun 04 - 08:31 & 4.670 & 3.736 & 5.5 & -306.46 & $g$ & 18.68 & 0.09 & 2.55 &1.40 & 140.25 & 206.79 \\
2019 Jun 06 - 06:37 & 4.668 & 3.725 & 5.1 & -304.60 & $r$ & 18.36 & 0.08 & 1.99 &1.77 & 116.28 & 171.48 \\
2019 Jun 06 - 08:33 & 4.668 & 3.724 & 5.1 & -304.53 & $g$ & 18.63 & 0.09 & 2.11 &1.39 & 143.63 & 211.82 \\
2019 Jun 07 - 09:33 & 4.668 & 3.719 & 4.9 & -303.52 & $r$ & 18.14 & 0.06 & 1.87 &1.39 & 140.87 & 207.78 \\
2019 Jun 08 - 06:36 & 4.667 & 3.714 & 4.8 & -302.67 & $r$ & 18.57 & 0.09 & 1.90 &1.73 & 94.15 & 138.88 \\
2019 Jun 08 - 09:12 & 4.667 & 3.713 & 4.8 & -302.57 & $g$ & 18.56 & 0.08 & 2.07 &1.37 & 150.52 & 222.03 \\
2019 Jun 09 - 06:36 & 4.667 & 3.709 & 4.6 & -301.71 & $r$ & 18.38 & 0.06 &1.89 &1.70 & 111.02 & 163.79 \\
2019 Jun 10 - 07:02 & 4.666 & 3.704 & 4.5 & -300.72 & $r$ & 18.34 & 0.06 & 1.90& 1.56 & 114.39 & 168.78 \\
2019 Jun 10 - 08:03 & 4.666 & 3.704 & 4.5 & -300.68 & $g$ & 18.70 & 0.10 & 2.27 &1.40 & 130.13 & 192.01 \\
2019 Jun 11 - 09:43 & 4.666 & 3.699 & 4.3 & -299.65 & $g$ & 18.90 & 0.10 & 1.93 &1.42 & 107.13 & 158.11 \\
2019 Jun 14 - 08:02 & 4.664 & 3.688 & 3.9 & -296.81 & $g$ & 18.74 & 0.22 & 2.03 &1.38 & 121.43 & 179.30 \\
2019 Jun 20 - 07:50 & 4.661 & 3.673 & 3.3 & -291.02 & $g$ & 18.61 & 0.22 & 2.51& 1.38 & 132.47 & 195.84 \\
2019 Jun 20 - 08:17 & 4.661 & 3.673 & 3.3 & -291.00 & $r$ & 18.15 & 0.10 & 1.97 &1.37 & 127.68 & 188.75 \\
2019 Jun 23 - 09:02 & 4.659 & 3.669 & 3.2 & -288.07 & $r$ & 18.23 & 0.08 & 2.41 &1.43 & 117.79 & 174.17 \\
2019 Jun 23 - 10:34 & 4.659 & 3.669 & 3.2 & -288.00 & $g$ & 18.77 & 0.18 &2.57& 1.84 & 113.53 & 167.87 \\
2019 Jun 26 - 07:37 & 4.657 & 3.667 & 3.2 & -285.22 & $r$ & 18.21 & 0.05 & 1.87 &1.37 & 119.75 & 177.06 \\
2019 Jun 26 - 08:02 & 4.657 & 3.667 & 3.2 & -285.20 & $g$ & 18.80 & 0.08 & 2.19 &1.38 & 110.22 & 162.98 \\
2019 Jul 01 - 06:45 & 4.655 & 3.671 & 3.5 & -280.41 & $r$ & 18.33 & 0.05 & 1.84 &1.40 & 108.62 & 160.51 \\
2019 Jul 01 - 06:46 & 4.655 & 3.671 & 3.5 & -280.41 & $r$ & 18.33 & 0.05 & 1.77 &1.39 & 108.62 & 160.51 \\
2019 Jul 01 - 07:44 & 4.655 & 3.671 & 3.5 & -280.37 & $r$ & 18.34 & 0.06 & 2.37 &1.38 & 107.63 & 159.04 \\
2019 Jul 01 - 07:44 & 4.655 & 3.671 & 3.5 & -280.37 & $r$ & 18.34 & 0.06 & 2.18 &1.38 & 107.63 & 159.04 \\
2019 Jul 03 - 07:10 & 4.654 & 3.674 & 3.8 & -278.45 & $r$ & 18.17 & 0.05 & 2.31 &1.37 & 127.49 & 188.29 \\
2019 Jul 03 - 07:32 & 4.654 & 3.674 & 3.8 & -278.44 & $g$ & 18.58 & 0.07 & 2.14& 1.37 & 138.51 & 204.56 \\
2019 Jul 04 - 07:36 & 4.653 & 3.676 & 3.9 & -277.46 & $r$ & 18.09 & 0.05 & 2.04& 1.38 & 137.86 & 203.57 \\
2019 Jul 06 - 06:43 & 4.652 & 3.681 & 4.2 & -275.56 & $g$ & 18.74 & 0.09 & 1.92& 1.38 & 121.73 & 179.68 \\
2019 Jul 08 - 07:32 & 4.651 & 3.687 & 4.5 & -273.59 & $g$ & 18.48 & 0.07 & 1.92 &1.39 & 156.89 & 231.49 \\
2019 Jul 09 - 05:29 & 4.651 & 3.690 & 4.6 & -272.70 & $r$ & 18.13 & 0.04 & 1.58& 1.48 & 137.39 & 202.69 \\
2019 Jul 09 - 07:36 & 4.651 & 3.690 & 4.6 & -272.61 & $g$ & 18.61 & 0.07 & 1.84& 1.40 & 139.94 & 206.46 \\
2019 Jul 10 - 06:01 & 4.650 & 3.694 & 4.8 & -271.71 & $r$ & 18.21 & 0.06 & 2.02& 1.41& 128.82 & 190.02 \\
2019 Jul 10 - 06:02 & 4.650 & 3.694 & 4.8 & -271.71 & $r$ & 18.06 & 0.06 & 2.30&1.41 & 147.90 & 218.17 \\
2019 Jul 10 - 06:13 & 4.650 & 3.694 & 4.8 & -271.70 & $r$ & 18.27 & 0.06 & 1.96 &1.39 & 121.89 & 179.80 \\
2019 Jul 10 - 06:31 & 4.650 & 3.694 & 4.8 & -271.69 & $r$ & 18.20 & 0.07 & 2.44& 1.38 & 130.01 & 191.77 \\
2019 Jul 10 - 06:41 & 4.650 & 3.694 & 4.8 & -271.68 & $r$ & 18.15 & 0.05 & 2.04& 1.37 & 136.14 & 200.81 \\
2019 Jul 10 - 06:42 & 4.650 & 3.694 & 4.8 & -271.68 & $r$ & 18.18 & 0.06 & 2.14 &1.37 & 132.43 & 195.34 \\
2019 Jul 11 - 07:29 & 4.649 & 3.698 & 4.9 & -270.68 & $g$ & 18.70 & 0.13 & 1.90 &1.40 & 130.73 & 192.82 \\
2019 Jul 12 - 06:32 & 4.649 & 3.702 & 5.1 & -269.74 & $r$ & 18.23 & 0.11 & 2.46& 1.37 & 128.40 & 189.36 \\
2019 Jul 12 - 06:33 & 4.649 & 3.702 & 5.1 & -269.74 & $r$ & 18.25 & 0.10 & 2.35 &1.37 & 126.06 & 185.90 \\
2019 Jul 12 - 07:24 & 4.649 & 3.702 & 5.1 & -269.71 & $g$ & 18.53 & 0.15 &2.01 &1.40 & 154.37 & 227.66 \\
2019 Jul 13 - 06:03 & 4.648 & 3.706 & 5.3 & -268.79 & $r$ & 18.12 & 0.10 & 2.49 &1.39 & 143.40 & 211.46 \\
2019 Jul 13 - 07:51 & 4.648 & 3.707 & 5.3 & -268.72 & $g$ & 18.80 & 0.24 & 1.69 &1.45 & 121.56 & 179.25 \\
2019 Jul 16 - 05:21 & 4.647 & 3.721 & 5.8 & -265.91 & $g$ & 18.57 & 0.28 & 2.10 &1.39 & 154.15 & 227.28 \\
2019 Jul 16 - 06:32 & 4.647 & 3.721 & 5.8 & -265.86 & $r$ & 18.26 & 0.16 & 1.92 &1.39 & 129.40 & 190.79 \\
2019 Jul 20 - 06:54 & 4.645 & 3.745 & 6.5 & -261.96 & $g$ & 18.65 & 0.16 & 2.67 &1.45 & 148.69 & 219.26 \\
2019 Jul 21 - 05:12 & 4.645 & 3.751 & 6.6 & -261.05 & $r$ & 18.22 & 0.04 & 1.52 &1.38 & 140.36 & 206.99 \\
2019 Jul 21 - 05:13 & 4.645 & 3.751 & 6.6 & -261.05 & $r$ & 18.18 & 0.04 & 1.58 &1.38 & 145.63 & 214.76 \\
2019 Jul 21 - 05:43 & 4.644 & 3.751 & 6.6 & -261.03 & $r$ & 18.17 & 0.05 & 1.44 &1.37 & 146.91 & 216.66 \\
2019 Jul 21 - 06:07 & 4.644 & 3.751 & 6.7 & -261.01 & $r$ & 18.34 & 0.05 & 1.46 &1.38 & 126.07 & 185.94 \\
2019 Jul 21 - 06:44 & 4.644 & 3.752 & 6.7 & -260.99 & $r$ & 18.35 & 0.09 & 2.67 &1.43 & 124.98 & 184.33 \\
2019 Jul 21 - 06:44 & 4.644 & 3.752 & 6.7 & -260.99 & $r$ & 18.31 & 0.08 & 2.71& 1.44 & 129.68 & 191.25 \\
2019 Jul 30 - 06:02 & 4.640 & 3.820 & 8.2 & -252.26 & $r$ & 18.21 & 0.05 & 1.85& 1.42 & 155.19 & 229.33 \\
2019 Jul 30 - 06:03 & 4.640 & 3.820 & 8.2 & -252.26 & $r$ & 18.20 & 0.04 & 1.70& 1.42 & 156.63 & 231.45 \\
2019 Jul 30 - 06:29 & 4.640 & 3.820 & 8.2 & -252.24 & $r$ & 18.27 & 0.05 & 1.92& 1.49 & 146.85 & 217.00 \\
2019 Jul 30 - 06:30 & 4.640 & 3.820 & 8.2 & -252.24 & $r$ & 18.30 & 0.05 & 1.82& 1.49 & 142.85 & 211.08 \\
2019 Aug 03 - 04:02 & 4.638 & 3.856 & 8.8 & -248.45 & $r$ & 18.31 & 0.10 & 1.65 &1.40 & 147.13 & 217.68 \\
2019 Aug 12 - 03:38 & 4.634 & 3.949 & 10.0 & -239.69 & $r$ & 18.11 & 0.16 & 3.23& 1.39 & 192.88 & 286.40 \\
2019 Aug 16 - 07:30 & 4.632 & 3.996 & 10.5 & -235.63 & $r$ & 18.34 & 0.16 &1.83 &1.78 & 162.34 & 241.48 \\
2019 Aug 19 - 03:27 & 4.631 & 4.030 & 10.8 & -232.87 & $r$ & 18.33 & 0.09 & 1.41 &1.38 & 168.23 & 250.52 \\
2019 Aug 23 - 04:03 & 4.629 & 4.080 & 11.2 & -228.94 & $g$ & 18.73 & 0.09 &1.80 &1.39 & 191.39 & 285.49 \\
2019 Aug 28 - 04:52 & 4.627 & 4.145 & 11.6 & -224.02 & $r$ & 18.34 & 0.06 & 1.73 &1.54 & 180.68 & 269.98 \\
2019 Aug 28 - 04:53 & 4.627 & 4.145 & 11.6 & -224.02 & $r$ & 18.37 & 0.06 & 1.78 &1.54 & 175.75 & 262.63 \\
2019 Aug 28 - 05:28 & 4.627 & 4.146 & 11.6 & -223.99 & $r$ & 18.43 & 0.06 & 1.56 &1.71 & 166.38 & 248.63 \\
2019 Aug 28 - 05:29 & 4.627 & 4.146 & 11.6 & -223.99 & $r$ & 18.51 & 0.06 & 1.68& 1.72 & 154.56 & 230.97 \\
2019 Aug 29 - 04:45 & 4.627 & 4.159 & 11.7 & -223.05 & $r$ & 18.37 & 0.06 & 2.22 &1.53& 177.51 & 265.38 \\
2019 Aug 29 - 04:46 & 4.627 & 4.159 & 11.7 & -223.05 & $r$ & 18.43 & 0.07 & 2.42 &1.53 & 167.97 & 251.11 \\
2019 Aug 29 - 05:46 & 4.627 & 4.159 & 11.7 & -223.00 & $r$ & 18.37 & 0.06 & 1.60 &1.86& 177.51 & 265.38 \\
2019 Aug 29 - 05:46 & 4.627 & 4.159 & 11.7 & -223.00 & $r$ & 18.38 & 0.06 & 1.62& 1.87 & 175.88 & 262.95 \\
2019 Aug 29 - 06:13 & 4.627 & 4.159 & 11.7 & -222.99 & $g$ & 18.85 & 0.14 & 1.96& 2.13 & 180.81 & 270.31 \\
2019 Sep 02 - 03:50 & 4.625 & 4.212 & 12.0 & -219.17 & $g$ & 18.94 & 0.11 & 1.84 &1.43& 172.19 & 257.81 \\
2019 Sep 02 - 05:06 & 4.625 & 4.213 & 12.0 & -219.12 & $r$ & 18.56 & 0.09 & 2.20 &1.70 & 154.25 & 230.94 \\
2019 Sep 03 - 03:40 & 4.624 & 4.226 & 12.0 & -218.20 & $g$ & 18.73 & 0.14 & 1.85 &1.41 & 210.23 & 314.76 \\
2019 Sep 09 - 03:04 & 4.622 & 4.309 & 12.3 & -212.35 & $g$ & 18.97 & 0.26 & 1.95 &1.40 & 176.75 & 265.04 \\
2019 Sep 09 - 05:38 & 4.622 & 4.311 & 12.3 & -212.25 & $r$ & 18.65 & 0.22 & 2.84& 2.23 & 149.88 & 224.75 \\
2019 Sep 25 - 03:09 & 4.616 & 4.540 & 12.5 & -196.68 & $r$ & 18.61 & 0.12 & 4.13& 1.57 & 173.11 & 259.86 \\
2019 Sep 25 - 04:55 & 4.616 & 4.542 & 12.5 & -196.61 & $g$ & 19.11 & 0.23 & 2.74 &2.55 & 173.26 & 260.09 \\
2019 Oct 01 - 03:35 & 4.614 & 4.628 & 12.4 & -190.78 & $g$ & 18.95 & 0.12 & 2.15 &1.67 & 207.61 & 311.48 \\
2019 Oct 09 - 02:21 & 4.611 & 4.742 & 12.2 & -182.98 & $g$ & 19.16 & 0.25 & 1.78& 1.49 & 178.27 & 267.18 \\
2019 Oct 12 - 02:42 & 4.610 & 4.785 & 12.0 & -180.02 & $r$ & 18.50 & 0.12 & 2.11 &1.62 & 208.92 & 312.79 \\
2019 Oct 16 - 02:42 & 4.608 & 4.840 & 11.8 & -176.09 & $g$ & 19.09 & 0.17 & 2.40 &1.70 & 195.32 & 292.15 \\
2019 Oct 19 - 02:09 & 4.607 & 4.881 & 11.6 & -173.17 & $r$ & 18.71 & 0.12 & 1.47 &1.59 & 176.65 & 263.96 \\
2019 Oct 19 - 02:36 & 4.607 & 4.882 & 11.6 & -173.15 & $g$ & 19.10 & 0.14 & 1.94 &1.74 & 195.56 & 292.23 \\
2019 Oct 22 - 02:40 & 4.606 & 4.922 & 11.4 & -170.20 & $g$ & 19.14 & 0.15 & 2.32 &1.852 & 190.28 & 284.07 \\
2019 Oct 26 - 02:30 & 4.605 & 4.975 & 11.1 & -166.27 & $g$ & 19.42 & 0.22 & 3.93 &1.891 & 148.68 & 221.69 \\
2019 Oct 26 - 02:35 & 4.605 & 4.975 & 11.1 & -166.27 & $r$ & 18.57 & 0.12 & 2.23 &1.935 & 205.24 & 306.01 \\
2019 Nov 08 - 01:48 & 4.601 & 5.135 & 9.9 & -153.51 & $r$ & 18.60 & 0.19 & 2.13 &1.755 & 204.05 & 302.88 \\
\hline
\caption{Columns: (1) observation date; (2) heliocentric distance; (3) topo-centric distance; (4) phase angle; (5) difference between time of observation $T$ and time of perihelion $T_p$; (6) filter; (7) 2$\times\,10^4$ km aperture mag; (8) 1-$\sigma$ mag uncertainty; (9) in-image seeing of observations; (10) airmass of observations; (11) \afroz of observations; (12) cross-section of observations}
\end{longtable}

\end{document}